\documentclass[onecolumn,aps,prd,preprintnumbers,showpacs,superscriptaddress,nofootinbib,amsmath,amssymb,floats,floatfix,showkeys,notitlepage,longbibliography]{revtex4-1}

\usepackage{comment}
\usepackage{graphicx}
\usepackage{subfigure}
\usepackage{palatino}
\usepackage{sans}
\usepackage[commandnameprefix=always]{changes}
\usepackage{xcolor}
\colorlet{BLUE}{blue}
\usepackage{hyperref}
\hypersetup{colorlinks=true,linkcolor=blue,urlcolor=blue,citecolor=blue}
\usepackage[toc,page]{appendix}
\usepackage[normalem]{ulem}
\usepackage{orcidlink}
\usepackage{adjustbox}
\usepackage{latexsym}
\usepackage{amsmath,amssymb,amsfonts}
\usepackage{dcolumn}
\usepackage{bm}
\usepackage{tikz}
\usepackage{bigints}
\usepackage{array,tabularx,multirow,booktabs}
\usepackage[tracking=true]{microtype}
\SetTracking{}{500}
\SetTracking{encoding={*}, shape=sc}{40}
\UseRawInputEncoding
\allowdisplaybreaks

\def\0{{\sst{(0)}}}
\def\1{{\sst{(1)}}}
\def\2{{\sst{(2)}}}
\def\3{{\sst{(3)}}}
\def\4{{\sst{(4)}}}
\def\5{{\sst{(5)}}}
\def\6{{\sst{(6)}}}
\def\7{{\sst{(7)}}}
\def\8{{\sst{(8)}}}
\def\sst#1{{\scriptscriptstyle #1}}

\begin{document} \sloppy

\title{Spinning Particle Dynamics and Observational Redshift around an Asymptotically Flat Symmergent Black Hole}

\author{Beyhan Puli{\c c}e
\orcidlink{0000-0002-5277-6906}
}
\email{beyhan.pulice@istinye.edu.tr}
\affiliation{Department of Basic Sciences, {\.I}stinye University, 34396, {\.I}stanbul, T\"{u}rkiye}

\author{Ali \"Ovg\"un
\orcidlink{0000-0002-9889-342X}
}
\email{ali.ovgun@emu.edu.tr}
\affiliation{Physics Department, Eastern Mediterranean
University, Famagusta, 99628 North Cyprus, via Mersin 10, Turkiye.}

\begin{abstract}

We investigate timelike particle dynamics, collision energetics, and photon frequency shifts in the perturbative variable-scalar curvature branch of asymptotically flat symmergent gravity. The low-energy vacuum action contains an $R^{2}$ correction whose coefficient is set by the boson--fermion imbalance of the underlying quantum field theory. At linear order, the exterior geometry is conformal to Schwarzschild spacetime through a radial mode satisfying a linear equation. We retain an independent boundary-condition-dependent amplitude and restrict the analysis to the perturbative domain. The two signs of the symmergent parameter $\gamma$ yield distinct profiles: $\gamma>0$ gives a Yukawa-suppressed deformation, whereas $\gamma<0$ produces an oscillatory inverse-radius deformation. We derive radial equations, effective potentials, circular-orbit and marginal-stability conditions for neutral, electrically charged, and spinning massive particles. Charged particles are treated in the test-field approximation, while spinning particles obey the Mathisson--Papapetrou--Dixon equations with the Tulczyjew condition. We also compute the center-of-mass energy of neutral-particle collisions and the frequency shifts of photons emitted tangentially by circular geodesic sources and detected by a static observer at infinity. The redshift and blueshift factors satisfy $(1+z_{+})(1+z_{-})=1/A(r_e)$, directly linking their product to the lapse function at emission. The $\gamma>0$ branch yields smooth, short-range deviations from Schwarzschild dynamics, whereas the $\gamma<0$ branch can generate oscillatory radial bands admitting circular-orbit solutions whose stability must be tested independently. These observables provide complementary probes of the variable-curvature sector, although their quantitative interpretation also depends on the deformation amplitude and, for the oscillatory branch, its phase.

\end{abstract}

\date{\today}



\maketitle

\section{Introduction}

Black holes provide one of the most sensitive arenas for testing gravitational dynamics beyond general relativity. Their near-horizon and strong-field regions control a wide range of observables, including particle motion, accretion dynamics, photon rings, black-hole shadows, gravitational-wave ringdown, and high-energy collision processes. The Event Horizon Telescope observations of M87* and Sgr A* have already demonstrated that horizon-scale measurements can be used to probe compact-object geometry in the strong-field regime \cite{EventHorizonTelescope:2019pgp,EventHorizonTelescope:2022wkp, Atamurotov:2013sca,Atamurotov:2015nra,Abdujabbarov:2016hnw,Allahyari:2019jqz,Vagnozzi:2019apd,Vagnozzi:2022moj}.

Among the many possibilities for extending general relativity, Symmergent gravity is particularly interesting because its gravitational sector is tied directly to the particle content of the underlying quantum field theory. In this framework, gravity emerges together with a quadratic curvature correction, and the coefficient of the $R^2$ term is fixed by the difference between the number of bosonic and fermionic degrees of freedom \cite{demir0,demir1,demir2,demir3,Demir2023}. Thus, deviations from Einstein gravity are not introduced as arbitrary phenomenological parameters, but are related to the microscopic field spectrum. This feature makes black hole solutions in symmergent gravity useful for connecting strong-gravity phenomenology with particle-content information beyond the standard model.

Static black holes in symmergent gravity have already been investigated in several contexts, including shadow observables, weak deflection of light, thermodynamic properties, and asymptotically flat variable-curvature configurations \cite{Pulice:2024wjg,Symmergent-bh,Symmergent-bh2,Symmergent-bh3,symmergentresults,Cimdiker2023,Pulice2023}. A key feature of the asymptotically flat branch is that constant-scalar-curvature configurations do not probe the quadratic curvature sector, whereas variable-scalar-curvature solutions do. In the latter case, the black hole geometry is conformally related to Schwarzschild spacetime, and the conformal factor is governed by a linear differential equation involving the Symmergent parameter $\gamma$. Since $\gamma$ is determined by the boson--fermion number difference, the geometry carries explicit information about the underlying particle spectrum.

The purpose of the present work is to study this variable-curvature asymptotically flat branch from the viewpoint of timelike probes. Instead of focusing on null geodesics, shadow radius, or weak deflection, we analyze the dynamics of massive particles in the Symmergent black hole background. This provides an independent and physically complementary test of the same geometry. Massive particle motion is especially useful because circular orbits, effective potentials, ISCO locations, and center-of-mass energies are sensitive not only to the lapse function but also to the full radial and angular structure of the metric. Such quantities may encode deviations that are not immediately visible in purely optical observables.

We consider three classes of probes. First, we study neutral massive particles following timelike geodesics and determine their effective potential, angular momentum, radial motion, ISCO structure, and collision energy. Second, we introduce charged test particles moving in a test electromagnetic potential and analyze how the combined effect of the Symmergent deformation and electromagnetic coupling modifies circular orbits and ISCO behavior. Third, we investigate spinning particles using the Mathisson--Papapetrou--Dixon (MPD) equations supplemented by the Tulczyjew spin-supplementary condition (SSC). In this case, spin--curvature coupling provides an additional channel through which the symmergent deformation can influence particle dynamics.

The dynamics of spinning test particles has recently become an active tool for probing strong-field gravity, because spin--curvature coupling modifies circular orbits, ISCO locations, epicyclic frequencies, and possible high-energy collision channels. In particular, spinning-particle motion has been analyzed in non-spherical and deformed compact-object geometries such as the $\gamma$ spacetime \cite{Toshmatov:2019bda}, around traversable wormholes \cite{Benavides-Gallego:2021lqn}, and in magnetized Schwarzschild black holes where the combined effect of spin and magnetic interaction leads to significant modifications of orbital motion \cite{Abdulxamidov:2023jfq}. More recent studies have extended this program to magnetized black holes in modified gravity, Sgr A$^\ast$ phenomenology, loop-quantum and effective-quantum black holes, Reissner--Nordstrom-like solutions, charged hairy black holes, asymptotically safe regular black holes, and Horndeski-inspired geometries \cite{Oteev:2025jvf,Uktamov:2025bth,Abdukayumova:2026zir,Umarov:2025wzm,Abdukayumova:2025ztr,Rakhimova:2024hzt,Mannobova:2025uqf,Turakhonov:2026lia}. Related analyses of particle motion and epicyclic frequencies have also shown that orbital observables can be used to constrain deviations from standard black-hole backgrounds through astrophysical data \cite{AbdelmalekBouzenada:2026zwu,Mustafa:2026hti}. These developments motivate the present study of spinning-particle dynamics in the asymptotically-flat Symmergent black-hole branch, where the spin--curvature coupling can act as an additional probe of the conformal deformation controlled by the microscopic Bose--Fermi imbalance.

In addition to orbital and collision observables, frequency shifts of photons emitted from matter moving around compact objects provide a direct probe of the underlying spacetime geometry. Gravitational redshift, Doppler boosting, and light-propagation effects have long been known to play a central role in the optical appearance and spectra of black holes \cite{Bardeen:1972fi,Cunningham:1972zz,Cunningham:1975zz,Luminet:1979nyg}, and have recently been formulated in a general static, spherically symmetric setting as an observational method for extracting metric information from redshift/blueshift data \cite{Martinez-Valera:2023guj}. For circular emitters around an asymptotically-flat Symmergent black hole, both the gravitational and kinematic contributions are modified by the conformal deformation $\varphi(r)$ and therefore by the particle-content parameter $\gamma$. This makes the redshift and blueshift of photons a useful complementary diagnostic to the ISCO radius, effective potential, and center-of-mass collision energy studied above. In particular, because the two branches of the theory exhibit qualitatively different large-distance behavior---a Yukawa-suppressed correction for $\gamma>0$ and an oscillatory inverse-radius tail for $\gamma<0$---the observed frequency shift can distinguish smooth Schwarzschild-like deviations from oscillatory radial signatures. Motivated by this, we now derive the redshift formula for photons emitted by circular geodesic sources and show how the product of the redshift and blueshift factors can be used to reconstruct the metric lapse function of the asymptotically-flat Symmergent black hole.

A central distinction in the analysis is the sign of the Symmergent parameter $\gamma$. For $\gamma>0$, corresponding to a fermion-dominated particle spectrum, the conformal deformation is Yukawa suppressed at large radius. For $\gamma < 0$, which corresponds to a boson-dominated spectrum, the deformation has an oscillatory inverse-radius tail. These two asymptotic behaviors lead to qualitatively different modifications of the effective potentials and orbital observables. We therefore treat both branches explicitly and compare their physical consequences.

The paper is organized as follows. In Sec.~\ref{sec2}, we summarize the symmergent gravity framework, define the particle-content parameter $\gamma$, and introduce the perturbative asymptotically flat variable-curvature exterior. In Sec.~\ref{sec3}, we analyze neutral, charged, and spinning particle motion, derive the corresponding effective potentials and circular-orbit conditions, and compute the center-of-mass collision energy for neutral particles. We then discuss the behavior of circular orbits and marginally stable circular orbits in the two branches of the theory. In Sec.~\ref{sec4}, we study observational frequency shifts from the asymptotically flat symmergent black-hole exterior. Finally, in Sec.~\ref{sec5}, we summarize our results and their domain of validity.

\section{Symmergent gravity and the black hole background}
\label{sec2}

Symmergent gravity is an emergent gravity framework in which gauge symmetry restoration in a cutoff quantum field theory induces, at low energies, an Einstein--Hilbert term supplemented by a quadratic curvature correction \cite{demir0,demir1,demir2,demir3,Demir2023}. In the vacuum sector relevant for the present work, the effective action may be written as
\begin{align}
S
=
-\frac{c_{\rm O}}{16}
\int d^4x\sqrt{-g}
\left(
R^2+6\gamma G_N^{-1}R
\right),
\label{fr-action}
\end{align}
where $G_N$ is Newton's constant. The coefficient of the $R^2$ term is fixed by the particle content,
\begin{align}
c_{\rm O}
=
\frac{n_B-n_F}{128\pi^2},
\qquad
\gamma
=
-\frac{1}{6\pi c_{\rm O}}
=
-\frac{64\pi}{3(n_B-n_F)} .
\label{gamma}
\end{align}
Thus $\gamma>0$ corresponds to a fermion-dominated spectrum, $n_B<n_F$, whereas $\gamma<0$ corresponds to a boson-dominated spectrum, $n_B>n_F$. In the Bose--Fermi balanced limit, $n_B=n_F$, the coefficient $c_O$ of the quadratic-curvature term vanishes.

Variation of \eqref{fr-action} gives
\begin{align}
\left(R+3\gamma G_N^{-1}\right)R_{\mu\nu}
-
\frac14
\left(R+6\gamma G_N^{-1}\right)Rg_{\mu\nu}
-
\left(
\nabla_\mu\nabla_\nu-g_{\mu\nu}\Box
\right)R
=
0 .
\label{eins-eqn}
\end{align}
Asymptotically-flat constant-curvature black holes do not probe the quadratic curvature sector, since their geometry is insensitive to $c_{\rm O}$. The variable scalar curvature branch is therefore the relevant one for testing the particle content dependence of symmergent gravity.

The asymptotically flat variable-curvature exterior used below is perturbatively conformal to Schwarzschild spacetime. We introduce an explicit dimensionless deformation amplitude $\epsilon$ and write
\begin{align}
\varphi(r)=\epsilon f(r),
\qquad
|\epsilon f(r)|\ll1.
\label{perturbative-varphi}
\end{align}
In the dimensionless variables $M\rightarrow G_N^{1/2}M\equiv M$ and $r\rightarrow G_N^{-1/2}r\equiv r$, the metric takes the compact form \cite{nguyen3}
\begin{align}
ds^2
&=
e^{-\varphi(r)}
\left[
-\Psi(r)dt^2
+
\frac{dr^2}{\Psi(r)}
+
r^2d\Omega^2
\right]
+
\mathcal O(\epsilon^2),
\nonumber\\
\Psi(r)
&=
1-\frac{2M}{r}.
\label{metric0}
\end{align}
The exponential notation is used only as compact bookkeeping; all physical predictions are restricted to first order in $\epsilon$.

The radial deformation function satisfies
\begin{align}
\left[
r^2\Psi(r)\varphi'(r)
\right]'
=
\gamma r^2\varphi \, .
\label{varphi-eq}
\end{align}

The validity condition is therefore
\begin{align}
|\varphi(r)|
=
|\epsilon f(r)|
\leq \eta\ll1
\label{smallphi}
\end{align}
throughout the complete radial interval used in each calculation. In the numerical analysis below, we adopt the conservative requirement $\eta\leq0.1$.
At large radius, $\Psi(r)\simeq1$, and \eqref{varphi-eq} reduces to
\begin{align}
\left(r^2\varphi'\right)'
\simeq
\gamma r^2\varphi .
\end{align}

For later use, we write the metric as
\begin{align}
ds^2
=
-A(r)dt^2
+
\frac{dr^2}{B(r)}
+
C(r)d\Omega^2,
\label{metric-antz}
\end{align}
with 
\begin{align}
A(r)=e^{-\varphi(r)}\Psi(r),
\qquad
B(r)=e^{\varphi(r)}\Psi(r),
\qquad
C(r)=e^{-\varphi(r)}r^2 .
\label{ABC-exact}
\end{align}
The conformal deformation changes the curvature invariants but does not regularize the central Schwarzschild singularity. Schematically,
\begin{align}
R_{\mu\nu\rho\sigma}R^{\mu\nu\rho\sigma}
=
\frac{48M^2}{r^6}
+
\delta K[\varphi,\varphi',\varphi'']
+
\mathcal{O}(\varphi^2)
\label{kretschmann-schematic}
\end{align}

At large radius, where $\Psi(r)\simeq1$, Eq.~\eqref{varphi-eq} reduces to
\begin{align}
\left(r^2\varphi'\right)'
\simeq
\gamma r^2\varphi .
\end{align}
The asymptotically decaying solutions may therefore be written as
\begin{align}
f(r)
\simeq
\begin{cases}
\dfrac{e^{-\sqrt{\gamma}\,r}}
{\sqrt{\gamma}\,r},
& \gamma>0,
\\[3mm]
\dfrac{\cos\!\left(\sqrt{|\gamma|}\,r+\delta\right)}
{\sqrt{|\gamma|}\,r},
& \gamma<0,
\end{cases}
\label{phi-soln}
\end{align}
$\delta$ is an integration phase in the oscillatory branch. We take it $\delta=0$ for the rest of the present work. Consequently, quantitative predictions cannot in general be expressed solely in terms of $n_B-n_F$ without specifying the boundary conditions that determine these integration constants.

\section{Particle Dynamics and Neutral-Particle Collisions near the Symmergent Black-Hole Exterior} \label{sec3}

\subsection{Neutral-Particle Dynamics and Collisions}

In this section, we analyze the neutral particle motion in the vicinity of the symmergent black hole. Given the spherically symmetric metric in (\ref{metric0}), it suffices to analyze timelike geodesics.  The geodesic Lagrangian for the massive neutral test particles is given by
\begin{equation}
\label{geodesic-eq_neutral}
L_{neutral} = \frac{1}{2} m g_{\mu\nu} \dot{x}^\mu \dot{x}^\nu,
\end{equation}
Dividing the Lagrangian by the particle mass and restricting the motion
to the equatorial plane, $\theta=\pi/2$, one obtains
\begin{align}
L_{neutral} = \frac{1}{2}\left( -A(r) \dot{t}^2 +\frac{\dot{r}^2}{B(r)}  + C(r) \dot{\phi}^2 \right).
\end{align}
which leads to the energy
\begin{equation}
    \mathcal{E} = A(r)\frac{dt}{d\lambda},
    \label{energy}
\end{equation}
and the angular momentum
\begin{equation}
   \mathcal{L} = C(r)\frac{d\phi}{d\lambda},
   \label{angular-mom}
\end{equation}
as two constants of motion with $\mathcal{E} = E/m$ and $\mathcal{L} = l/m$. Using the normalization condition $g_{\mu\nu} \dot{x}^\mu \dot{x}^\nu = -1$, one can obtain the equation of motion

\begin{align}
\dot{r}^2 \frac{A(r)}{B(r)} + V_{eff} = \mathcal{E}^2   
\end{align}
where
\begin{align}
V_{eff} =  A(r) \left ( 1 + \frac{\mathcal{L}^2}{C(r)} \right ) \,.
\label{eff-pot-neutral}
\end{align}

The circular motion of the particles is given by
\begin{align}
\mathcal{E}^2 = V_{eff} \,\, \text{with} \,\, \dot{r} = 0
\end{align}

For a timelike circular geodesic at radius $r=r_c$, the circularity conditions are
\begin{align}
\dot r(r_c)=0,
\qquad
\left.
\frac{\partial V_{\rm eff}}{\partial r}
\right|_{r=r_c}=0.
\label{neutral-circularity}
\end{align}
Solving these equations for the specific energy and specific angular momentum gives
\begin{align}
\mathcal E^2
&=
\frac{A(r)^2C'(r)}
{A(r)C'(r)-C(r)A'(r)},
\label{neutral-circular-energy}
\\
\mathcal L^2
&=
\frac{C(r)^2A'(r)}
{A(r)C'(r)-C(r)A'(r)}.
\label{neutral-circular-angular}
\end{align}
Physical future-directed circular timelike orbits must satisfy
\begin{align}
A(r)>0,
\qquad
\mathcal E^2>0,
\qquad
\mathcal L^2\geq0,
\qquad
A(r)C'(r)-C(r)A'(r)>0.
\label{neutral-circular-conditions}
\end{align}
The marginally stable circular orbit is determined by
\begin{align}
\left.
\frac{\partial^2V_{\rm eff}}{\partial r^2}
\right|_{r=r_{\rm ms}}=0,
\label{neutral-marginal-stability}
\end{align}
together with Eq.~\eqref{neutral-circularity}. Equivalently, along a regular family of circular timelike orbits one may solve
\begin{align}
\frac{d}{dr}
\left[
\frac{C(r)^2A'(r)}
{A(r)C'(r)-C(r)A'(r)}
\right]=0.
\label{neutral-isco-equation}
\end{align}
When several roots occur, the astrophysically relevant ISCO is the innermost boundary of the stable circular-orbit branch continuously connected to the stable large-radius region. The condition $r>3M$ is necessary in the present conformal geometry but is not sufficient to identify the physical ISCO.

\begin{figure}[h!]
   \centering
    \includegraphics[width=0.47\textwidth]{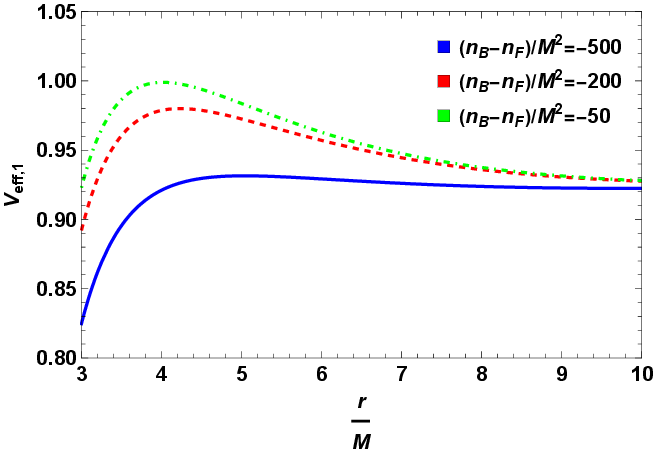}
    \includegraphics[width=0.465\textwidth]{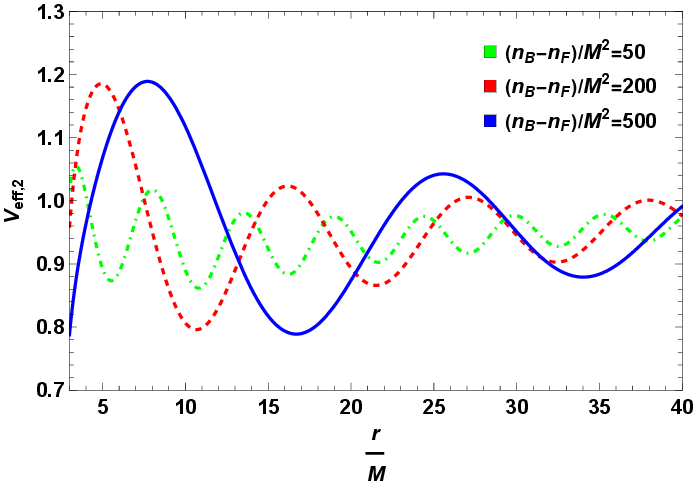}
    \caption{The variation of $V_{eff}$ of neutral particles with radius  for $\mathcal{L}=4 M$ and for selected negative values of $n_B - n_F$ ($\gamma >0$) (left panel) and positive values of $n_B - n_F$ ($\gamma < 0$) (right panel).}
    \label{Veff-neutral}
\end{figure}

Fig. \ref{Veff-neutral} illustrates how the effective potential for neutral test particles responds to the sign and magnitude of the Symmergent parameters namely $n_B - n_F$. For $n_B - n_F$ ($\gamma >0$)(left panel) the deviation induced by the Symmergent sector is localized to the strong-field region and becomes rapidly suppressed as $r/M$ increases, leading to a smooth, monotonic radial profile at large distances. For $n_B - n_F$ ($\gamma < 0$)(right panel), the potential develops oscillatory modulations whose amplitude and persistence increase with $\left| n_B - n_F \right|$, reflecting the oscillatory large-r behavior of the conformal sector in this branch.

\begin{figure}[h!]
   \centering
    \includegraphics[width=0.462\textwidth]{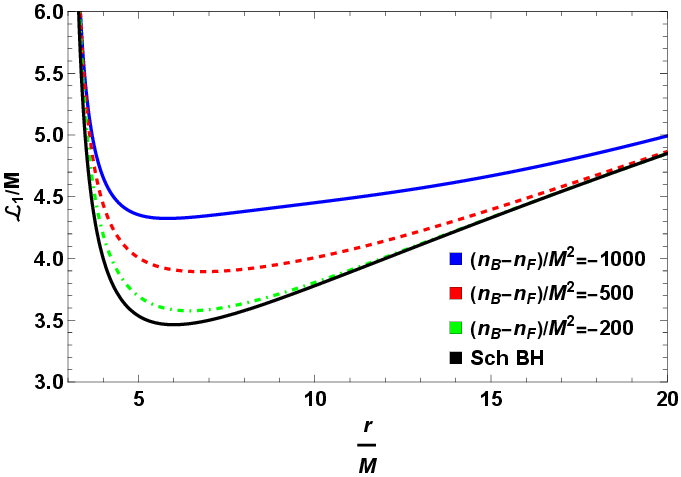}
    \includegraphics[width=0.464\textwidth]{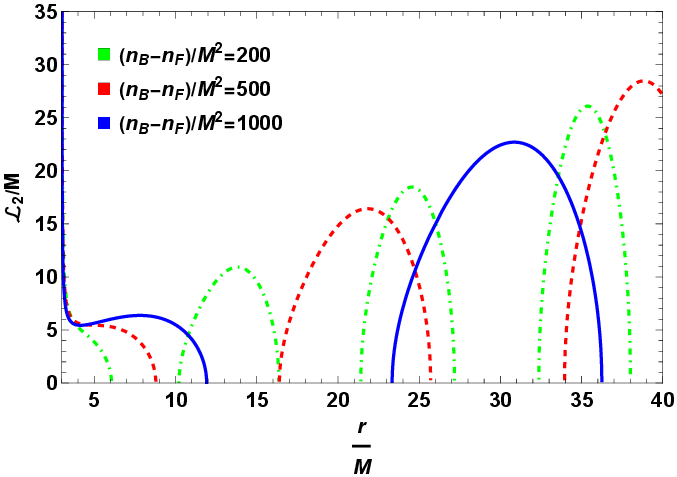}
    \caption{Specific angular momentum $\mathcal L$ of neutral massive particles on circular timelike geodesics as a function of radius. The left panel shows selected negative values of $n_B-n_F$ $(\gamma>0)$, whereas the right panel shows selected positive values of $n_B-n_F$ $(\gamma<0)$. Only radii satisfying Eq.~\eqref{neutral-circular-conditions} are physically admissible.}
    \label{Ang-mom-neutral}
\end{figure}

Fig.~\ref{Ang-mom-neutral} displays the specific angular momentum of neutral massive particles on circular timelike geodesics as a function of radius. In the symmergent background, the deviation from the Schwarzschild geometry is encoded in the conformal factor $\varphi(r)$, which deforms the metric potentials. For negative values of $n_B - n_F$ ($\gamma >0$)(left panel), the conformal deformation is Yukawa-suppressed at large radii (\ref{phi-soln}), so the angular momentum profile is mainly altered in the strong field region and becomes smooth as $r/M$ increases. By contrast, for positive values of $n_B - n_F$ ($\gamma < 0$)(right panel), the oscillatory $1/r$ tail of $\varphi(r)$ induces oscillatory modulations in $\mathcal{L}(r)$, reflecting the fact that the conformal sector can imprint additional radial structure on the circular orbit conditions.

\begin{figure}[h!]
   \centering
    \includegraphics[width=0.48\textwidth]{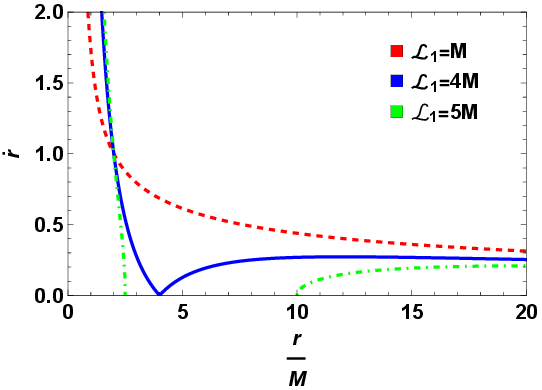}
    \includegraphics[width=0.48\textwidth]{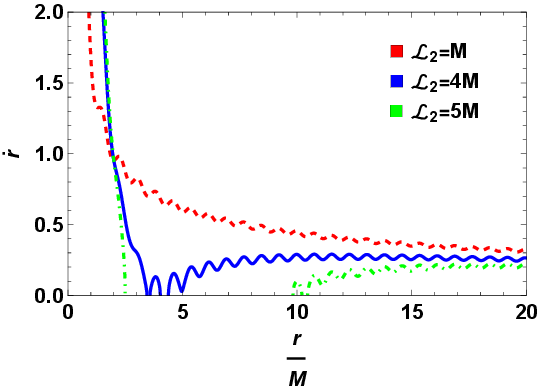}
    \caption{Radial velocity $\dot r$ as a function of the radius for fixed specific energy $\mathcal{E}=1$ and different angular momenta. The left panel corresponds to $n_B-n_F=-1$ $(\gamma>0)$, while the right panel corresponds to $n_B-n_F=1$ $(\gamma<0)$. Regions where the argument of the square root becomes negative are kinematically forbidden.}
    \label{rdot-neutral}
\end{figure}

Fig.~\ref{rdot-neutral} illustrates the radial velocity for representative values of the angular momentum. The condition $\dot r\geq0$ defines the kinematically accessible region, whereas $\dot r=0$ marks a turning point. In the $n_B-n_F<0$ $(\gamma>0)$ branch, the radial profile is smoothly deformed in the strong-field region. In the $n_B-n_F>0$ $(\gamma<0)$ branch, the oscillatory conformal tail can introduce additional turning points. Every displayed point must also satisfy $|\epsilon f(r)|\leq\eta$.

Here, we consider the collision of two neutral particles near the Symmergent black hole at some radial coordinate $r$. The center-of-mass energy of the two-particle system is defined by
\begin{align}
E_{\rm CM}^2
=
-\left(p_1^\mu+p_2^\mu\right)
\left(p_{1\mu}+p_{2\mu}\right).
\end{align}
Using $p_i^\mu=m_i v_i^\mu$, one obtains
\begin{align}
\frac{E_{\rm CM}^2}{2m_1m_2}
=
1-g_{\mu\nu}v_1^\mu v_2^\nu
+
\frac{(m_1-m_2)^2}{2m_1m_2}.
\end{align}
For equatorial motion in the metric \eqref{metric-antz}, this gives
\begin{align}
\frac{E_{\rm CM}^2}{2m_1m_2}
&=
1
+
\frac{\mathcal E_1\mathcal E_2}{A(r)}
-
\frac{\mathcal L_1\mathcal L_2}{C(r)}
-
\frac{\sigma_1\sigma_2}{B(r)}
\sqrt{\mathcal R_1(r)}
\sqrt{\mathcal R_2(r)}
+
\frac{(m_1-m_2)^2}{2m_1m_2},
\end{align}
where
\begin{align}
\mathcal R_i(r)
=
\frac{B(r)}{A(r)}\mathcal E_i^2
-
B(r)
\left(
1+\frac{\mathcal L_i^2}{C(r)}
\right),
\qquad
\sigma_i=\pm1 .
\end{align}
Here $\sigma_i=+1$ and $\sigma_i=-1$ distinguish outgoing and ingoing radial
motion, respectively. Thus $\sigma_1\sigma_2=+1$ corresponds to particles moving
in the same radial direction, whereas $\sigma_1\sigma_2=-1$ corresponds to a
head-on collision.

\begin{figure}[htb!]
   \centering
    \includegraphics[width=0.469\textwidth]{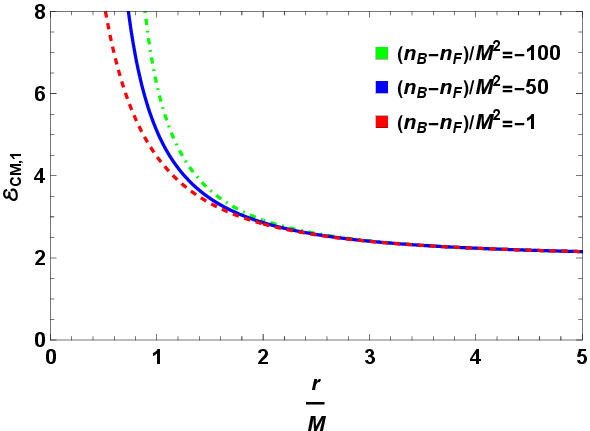}
    \includegraphics[width=0.48\textwidth]{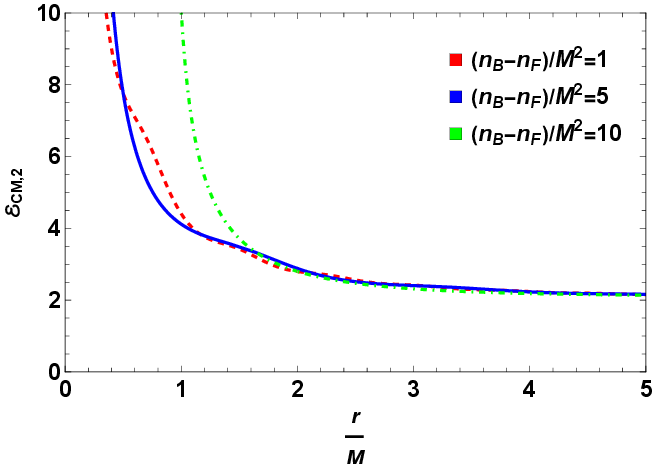}
    \caption{Center-of-mass energy of two neutral particles as a function of radius for $\mathcal L_1=-\mathcal L_2=2M$, $\mathcal{E}_1=\mathcal{E}_2=1 $ and $m_1=m_2=1$. Only radii satisfying $\mathcal R_1\geq0$, $\mathcal R_2\geq0$, and $|\epsilon f|\leq\eta$ are physical.}
    \label{COM-energy-neutral}
\end{figure}

Fig. \ref{COM-energy-neutral} shows the center-of-mass energy of two neutral particles. A collision at a given radius is physical only if both particles can reach that radius, namely if $\mathcal R_1(r)\geq0$ and $\mathcal R_2(r)\geq0$. The signs $\sigma_1$ and $\sigma_2$ must also be specified because particles moving in the same radial direction and particles undergoing a head-on collision have different center-of-mass energies. In the $\gamma>0$ branch, the perturbative correction is short-ranged, whereas in the $\gamma<0$ branch the oscillatory deformation may produce additional radial structure. All curves must be restricted to $\varphi(r) \leq \eta$.

\begin{figure}[htb!]
   \centering
   \includegraphics[width=0.44\textwidth]{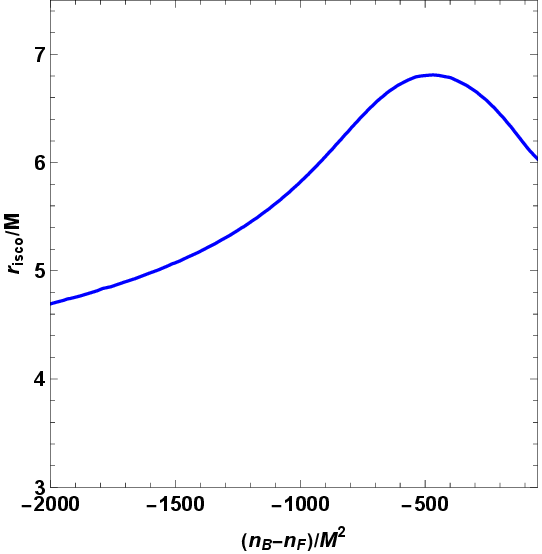}
     \includegraphics[width=0.465\textwidth]{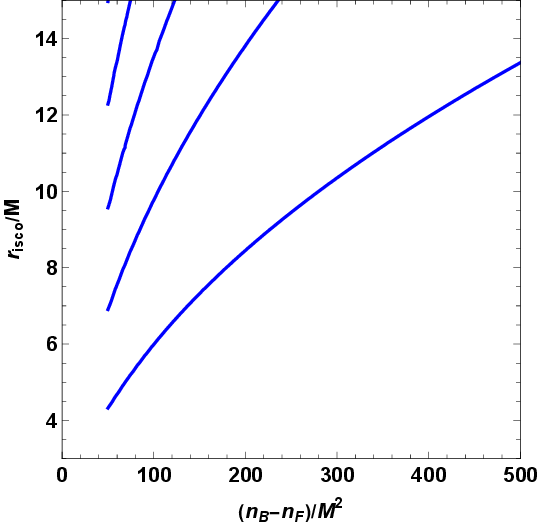}
    \caption{The variation of the ISCO radius of neutral particles with negative values of $n_B - n_F$ ($\gamma >0$)(left panel) and positive values of $n_B - n_F$ ($\gamma < 0$)(right panel).}
    \label{rISCO-neutral}
\end{figure}

Fig. \ref{rISCO-neutral} displays the dependence of the neutral particle ISCO radius on the symmergent parameter. The two branches exhibit qualitatively different behavior. For $n_B-n_F < 0$ $(\gamma > 0)$, the ISCO radius follows a smooth but non-monotonic profile. It remains outside the photon sphere and approaches the Schwarzschild scale $(r_{\rm ISCO}=6M)$ in the weak-deformation regime, while reaching a maximum at an intermediate value of $((n_B-n_F)/M^2)$. This indicates that the conformal symmergent correction shifts the marginally stable orbit most efficiently in an intermediate parameter range. For $n_B-n_F > 0$ $(\gamma < 0)$, the oscillatory conformal sector leads to several marginal-stability branches. This multi-branch structure reflects the trigonometric modulation of the metric functions and their derivatives, which can generate multiple radial roots of the ISCO condition. The condition $r_{\rm ms}>r_{\rm ps}=3M$ is necessary but not sufficient to identify the physical ISCO. For every marginal-stability root, we additionally require $\mathcal E_c^2>0$, $\mathcal L_c^2\geq0$, future-directed motion, and stability on the outer side of the root. When several disconnected stable intervals occur, the physical ISCO is the innermost boundary of the stable circular-orbit branch continuously connected to spatial infinity. Other marginal-stability roots correspond to additional stability boundaries and should not all be labeled as the ISCO.

\begin{figure}[htb!]
   \centering
    \includegraphics[width=0.46\textwidth]{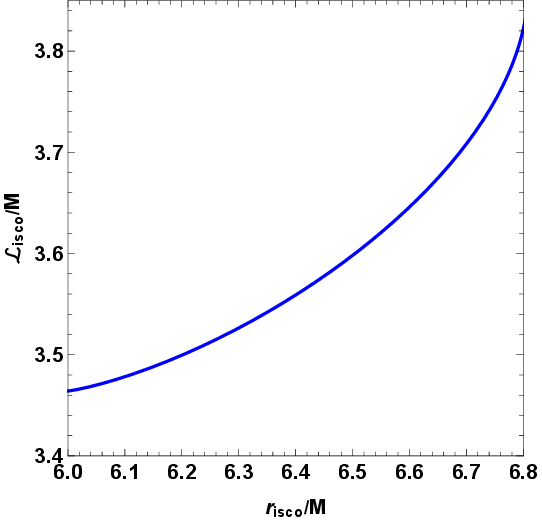}
     \includegraphics[width=0.46\textwidth]{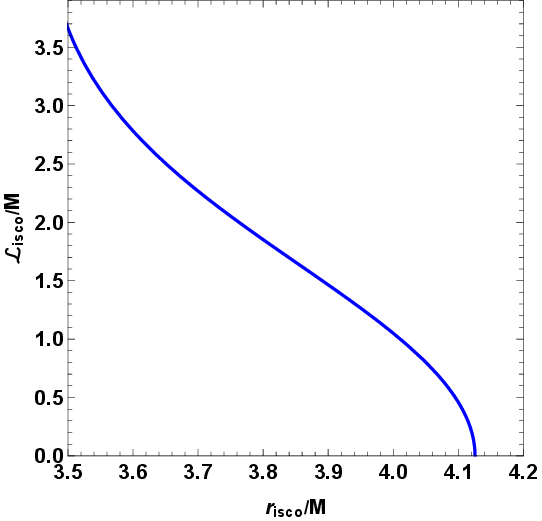}
    \caption{The variation of the angular momentum with  ISCO radius of neutral particles for negative values of $n_B - n_F$ ($\gamma >0$) (left panel) and positive values of $n_B - n_F$ ($\gamma < 0$) (right panel).}
    \label{ISCO-ang-mom-neutral}
\end{figure}

Fig. \ref{ISCO-ang-mom-neutral} shows the variation of the angular momentum at the ISCO for neutral massive particles as a function of the Bose-Fermi imbalance parameter. Since the symmergent parameter is fixed by  $n_B-n_F$, changing this quantity modifies the conformal deformation $\varphi(r)$ and hence the metric functions entering the circular-orbit and marginal-stability conditions. Therefore, $\mathcal L_{\rm isco}$ provides a direct measure of how the particle content parameter is imprinted on strong field orbital dynamics. In the $n_B-n_F<0$ branch $(\gamma>0)$, $\mathcal L_{\rm isco}$ decreases smoothly as $n_B-n_F$ approaches zero from below. This indicates that the angular momentum threshold required to support the innermost stable orbit becomes smaller as the corresponding conformal deformation changes across this parameter range. The smooth behavior is consistent with the non-oscillatory character of this branch. In the $n_B-n_F>0$ branch $(\gamma<0)$, $\mathcal L_{\rm isco}$ also decreases with increasing $n_B-n_F$, but over a much wider range and with a sharper drop near the upper end of the displayed interval. This stronger sensitivity reflects the oscillatory conformal sector, whose contribution to the derivatives of the metric functions can significantly modify the stability condition. Thus, the figure shows that the Bose-Fermi imbalance affects not only the radial location of the ISCO, but also the angular momentum scale required for stable timelike circular motion.

\begin{figure}[htb!]
   \centering
    \includegraphics[width=0.46\textwidth]{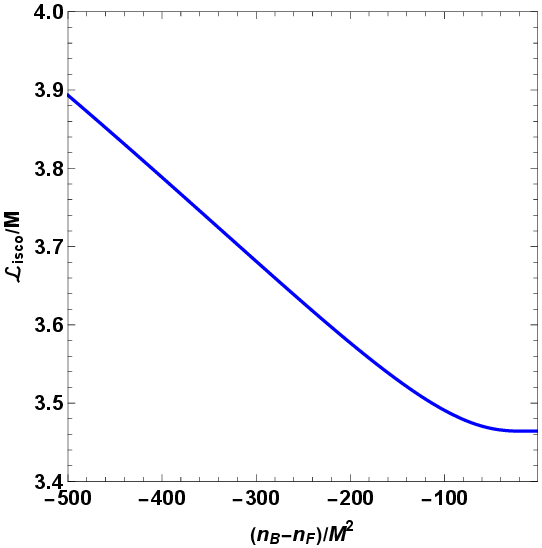}
     \includegraphics[width=0.445\textwidth]{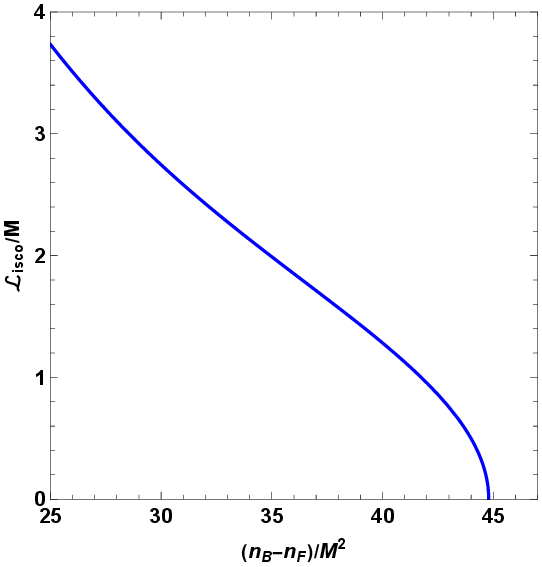}
    \caption{The variation of the angular momentum of the neutral particles at the ISCO for negative values of $n_B - n_F$ ($\gamma >0$) (left panel) and positive values of $n_B - n_F$ ($\gamma < 0$) (right panel).}
    \label{ISCO-ang-mom-neutral-2}
\end{figure}

Fig. \ref{ISCO-ang-mom-neutral-2} shows the relation between the ISCO radius and the corresponding angular momentum required for neutral massive particles to maintain marginally stable circular motion. This plot provides a compact characterization of how the symmergent conformal deformation modifies the strong-field orbital structure. Since both $r_{\rm isco}$ and $\mathcal L_{\rm isco}$ are determined by the circularity and stability conditions, their correlation reflects the combined effect of the metric functions $A(r)$ and $C(r)$ and their radial derivatives. In the $n_B-n_F<0$ branch $(\gamma>0)$, $\mathcal L_{\rm isco}$ increases monotonically with $r_{\rm isco}$. Thus, when the ISCO is shifted outward, a larger angular momentum is required to support the stable orbit. This behavior is consistent with a smooth deformation of the Schwarzschild-like orbital structure. In the $n_B-n_F>0$ branch $(\gamma<0)$, the trend is reversed: $\mathcal L_{\rm isco}$ decreases as $r_{\rm isco}$ increases. This indicates that the oscillatory conformal sector modifies the stability condition in a qualitatively different way, so that a larger ISCO radius does not necessarily correspond to a larger angular momentum threshold. The opposite slopes in the two panels therefore show that the sign of the Bose-Fermi imbalance affects not only the ISCO position itself, but also the angular momentum scale associated with the ISCO.

\begin{figure}[htb!]
   \centering
    \includegraphics[width=0.47\textwidth]{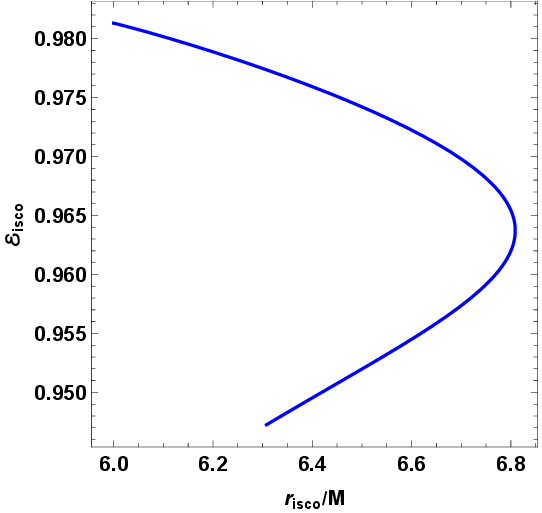}
     \includegraphics[width=0.45\textwidth]{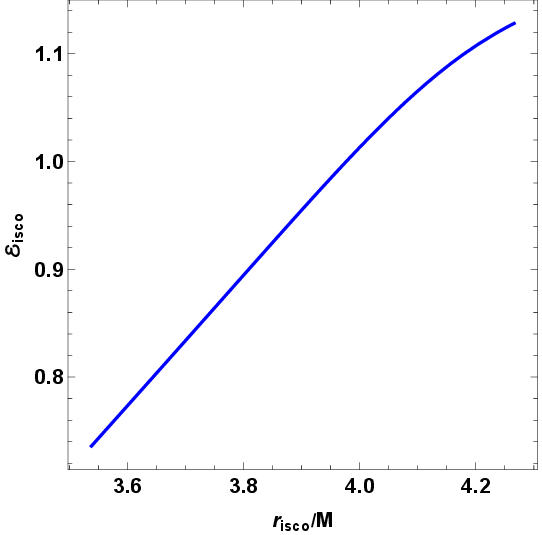}
    \caption{The relation between the conserved energy and ISCO radius for the neutral particles at the ISCO for negative values of $n_B - n_F$ ($\gamma >0$) (left panel) and positive values of $n_B - n_F$ ($\gamma < 0$) (right panel).}
    \label{ISCO-energy-neutral}
\end{figure}

Fig. \ref{ISCO-energy-neutral} shows the conserved energy evaluated at the ISCO as a function of the ISCO radius for neutral particles. In the $n_B-n_F$ branch $(\gamma>0)$, $\mathcal E_{\rm isco}$ remains below unity throughout the displayed range, indicating bound stable circular configurations. The curve is non-monotonic: the same ISCO radius can correspond to different energy values along the branch. This reflects the fact that $\mathcal E_{\rm isco}$ is not determined by $r_{\rm isco}$ alone, but also by the angular momentum threshold and by the symmergent deformation of the metric functions evaluated at the ISCO. In the $n_B-n_F > 0$ branch $(\gamma<0)$, shown in the right panel, $\mathcal E_{\rm isco}$ increases monotonically with $r_{\rm isco}$ and crosses the bound orbit threshold $\mathcal E_{\rm isco}=1$. Thus, part of the displayed branch corresponds to bound marginally stable orbits, whereas larger $r_{\rm isco}$ values require energy above the rest-mass threshold. This behavior indicates that the oscillatory conformal sector can significantly enhance the energetic cost of neutral particle stability. The contrast between the two panels shows that the sign of the Bose-Fermi imbalance changes not only the ISCO radius and angular momentum, but also the conserved energy associated with the ISCO.

\begin{figure}[h!]
   \centering
   \includegraphics[width=0.48\textwidth]{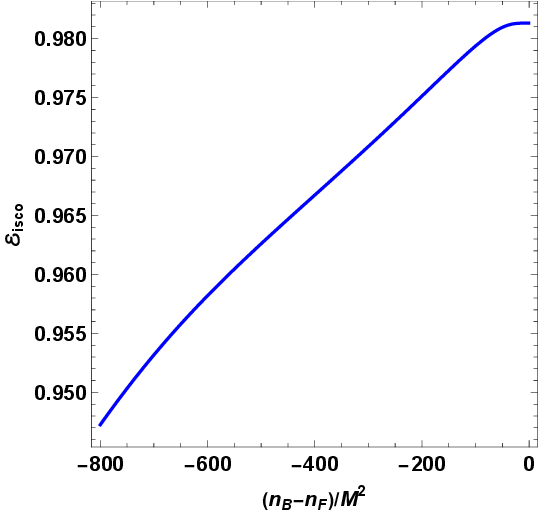}
    \includegraphics[width=0.47\textwidth]{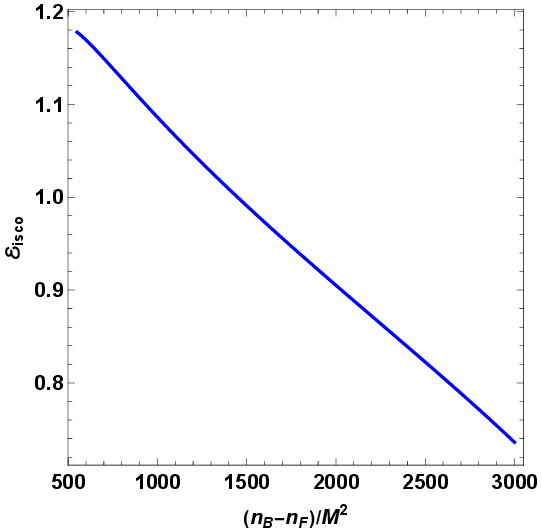}
    \caption{The variation of the energy at the ISCO of neutral particles with negative values of $n_B - n_F < 0$ ($\gamma >0$) (left panel) and positive values of $n_B - n_F > 0$ ($\gamma < 0$) (right panel).}
    \label{energy-ISCO-nM-neutral}
\end{figure}

Fig. \ref{energy-ISCO-nM-neutral} shows the dependence of the conserved specific energy at the ISCO, $\mathcal E_{\rm isco}$, on the Bose-Fermi imbalance parameter for neutral particles. In the $n_B-n_F < 0$ branch $(\gamma>0)$, $r_{\rm isco}$ increases smoothly as $n_B-n_F$ approaches zero from below. Throughout this branch, the energy remains below unity, indicating bound stable circular configurations. This smooth behavior is consistent with the non-oscillatory character of the conformal deformation in the $\gamma>0$ branch. In the $n_B-n_F > 0$ branch $(\gamma<0)$, $\mathcal E_{\rm isco}$ decreases monotonically as the Bose-Fermi imbalance increases. The curve crosses the threshold $\mathcal E_{\rm isco}=1$, so the lower $n_B-n_F$ part of the branch corresponds to energetically unbound marginal configurations, whereas the larger $n_B-n_F$ region corresponds to bound ISCO configurations. This indicates that the oscillatory conformal sector can substantially reduce the energy required for stable neutral circular motion. Overall, the two panels show that the sign of the Bose-Fermi imbalance qualitatively changes the energetic character of neutral particle ISCO configurations.

\begin{figure}[h!]
   \centering
   \includegraphics[width=0.47\textwidth]{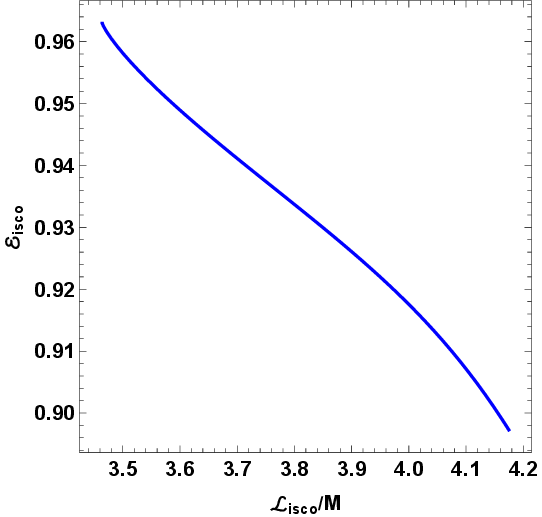}
    \includegraphics[width=0.455
    \textwidth]{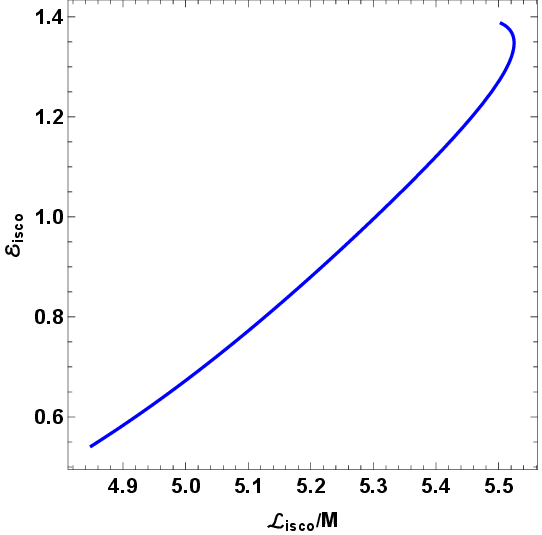}
    \caption{The relation between the specific energy $\mathcal E_{\rm isco}$ and the angular momentum $\mathcal L_{\rm isco}$ at the ISCO for neutral particles for negative values of $n_B - n_F < 0$ ($\gamma >0$) (left panel) and positive values of $n_B - n_F > 0$ ($\gamma < 0$) (right panel).}
    \label{E-L-ISCO-neutral}
\end{figure}

Fig. \ref{E-L-ISCO-neutral} shows the relation between the conserved specific energy $\mathcal E_{\rm isco}$ and the angular momentum $\mathcal L_{\rm isco}/M$ evaluated at the ISCO for neutral particles. This plot provides a phase space characterization of the stable circular orbit: instead of showing the ISCO radius directly, it displays how the energy cost of the orbit changes with the angular momentum threshold. In the $n_B-n_F < 0$ branch $(\gamma>0)$, $\mathcal E_{\rm isco}$ decreases as $\mathcal L_{\rm isco}/M$ increases. The entire curve remains below unity, indicating bound stable circular configurations. This inverse relation shows that, along this branch, larger angular momentum at the ISCO is associated with a lower conserved energy. Such behavior reflects the way the smooth conformal deformation modifies the balance between gravitational binding and centrifugal support. In the $n_B-n_F >0$ branch $(\gamma<0)$, the behavior is qualitatively different. Here $\mathcal E_{\rm isco}$ increases with $\mathcal L_{\rm isco}/M$, and the curve crosses the threshold $\mathcal E_{\rm isco}=1$. Therefore, the lower part of the branch corresponds to bound ISCO configurations, whereas the upper part corresponds to energetically unbound configurations. This demonstrates that the oscillatory conformal sector can strongly enhance the energy required for stable neutral circular motion. The opposite trends in the two panels show that the sign of the Bose-Fermi imbalance changes the energy-angular momentum structure of neutral particle ISCOs. Thus, $\mathcal E_{\rm isco}$-$\mathcal L_{\rm isco}/M$ provides a useful diagnostic of how the symmergent deformation reorganizes strong field timelike orbital stability.

\subsection{Dynamics of Charged Particles}

We now consider the motion of massive charged test particles in the same
background geometry. We treat the electromagnetic field as a test field, so that
it affects the particle motion but does not backreact on the metric. The charged
particle Lagrangian is taken as
\begin{align}
\label{geodesic-eq_charged}
L_{\rm charged}
=
\frac{1}{2}g_{\mu\nu}\dot{x}^{\mu}\dot{x}^{\nu}
+
q A_\mu \dot{x}^{\mu},
\end{align}
where $q$ is the charge-to-mass ratio of the particle. In the equatorial plane
$\theta=\pi/2$, and for a purely electric potential
$A_\mu=(A_t(r),0,0,0)$, this becomes
\begin{align}
L_{\rm charged}
=
\frac{1}{2}
\left[
-A(r)\dot t^2+\frac{\dot r^2}{B(r)}+C(r)\dot\phi^2
\right]
+
q A_t(r)\dot t .
\end{align}
The canonical time momentum is
\begin{align}
p_t
=
-A(r)\dot t+qA_t(r).
\end{align}
We define the conserved specific energy in the standard way,
$\mathcal E=-p_t$, so that
\begin{align}
\label{energy-charged}
\mathcal E
=
A(r)\dot t-qA_t(r),
\qquad
\dot t
=
\frac{\mathcal E+qA_t(r)}{A(r)}.
\end{align}
The conserved specific angular momentum is
\begin{align}
\mathcal L
=
C(r)\dot\phi .
\end{align}
Using $g_{\mu\nu}\dot x^\mu\dot x^\nu=-1$, the radial equation becomes
\begin{align}
\dot r^2
=
\frac{B(r)}{A(r)}
\left[\mathcal E+qA_t(r)\right]^2
-
B(r)
\left(
1+\frac{\mathcal L^2}{C(r)}
\right).
\label{rdot-charged-corrected}
\end{align}
The turning point condition $\dot r=0$ gives the two branches of the effective
potential,
\begin{align}
\mathcal E
=
V_{\rm eff}^{\pm}(r),
\qquad
V_{\rm eff}^{\pm}(r)
=
-qA_t(r)
\pm
\sqrt{
A(r)
\left(
1+\frac{\mathcal L^2}{C(r)}
\right)
}.
\label{eff-pot-charged-corrected}
\end{align}
In the following we use the positive-energy branch $V_{\rm eff}^{+}$. For circular orbits one imposes
\begin{align}
\dot r=0,
\qquad
\partial_r V_{\rm eff}^{+}=0.
\end{align}
Solving $\partial_r V_{\rm eff}^{+}=0$ for $\mathcal L^2$, one obtains
\begin{align}
\mathcal L_\pm^2
&=
\frac{1}
{\left[C(r)A'(r)-A(r)C'(r)\right]^2}
\Bigg\{
A(r)C(r)^2A'(r)C'(r)
-
C(r)^3
\left[
A'(r)^2
-
2q^2A(r)A_t'(r)^2
\right]
\nonumber\\
&\hspace{2.0cm}
\pm
2qA(r)C(r)^2A_t'(r)
\sqrt{
q^2C(r)^2A_t'(r)^2
-
C(r)A'(r)C'(r)
+
A(r)C'(r)^2
}
\Bigg\}.
\label{charged-angular-momentum-corrected}
\end{align}
Therefore the reality condition for circular charged-particle motion is
\begin{align}
q^2C(r)^2A_t'(r)^2
-
C(r)A'(r)C'(r)
+
A(r)C'(r)^2
\geq 0 .
\end{align}

The discriminant condition above is necessary but not sufficient because Eq.~\eqref{charged-angular-momentum-corrected} is obtained after algebraic squaring. Each candidate branch $\mathcal L_\pm^2$ must be substituted back into the original unsquared circularity equation
\begin{align}
0
&=
-qA_t'(r)
+
\frac{
A'(r)\left(1+\dfrac{\mathcal L^2}{C(r)}\right)
-
A(r)\dfrac{\mathcal L^2C'(r)}{C(r)^2}
}
{
2\sqrt{A(r)\left(1+\dfrac{\mathcal L^2}{C(r)}\right)}
}.
\label{charged-unsquared-condition}
\end{align}
A physical charged-particle circular orbit must additionally satisfy
\begin{align}
\mathcal L^2&\geq0,
&
A(r)&>0,
&
\mathcal E+qA_t(r)&>0,
&
\dot t&>0.
\label{charged-physical-conditions}
\end{align}
Candidate solutions that fail Eq.~\eqref{charged-unsquared-condition} or Eq.~\eqref{charged-physical-conditions} must be discarded as extraneous algebraic roots.

In the neutral limit $q\to0$, the two branches coincide and reduce to
\begin{align}
\mathcal L_+^2
=
\mathcal L_-^2
=
\frac{C(r)^2A'(r)}
{A(r)C'(r)-C(r)A'(r)} .
\end{align}
This is the standard circular-orbit expression for a neutral particle in a static,
spherically symmetric spacetime.

\begin{figure}[h!]
   \centering
    \includegraphics[width=0.47\textwidth]{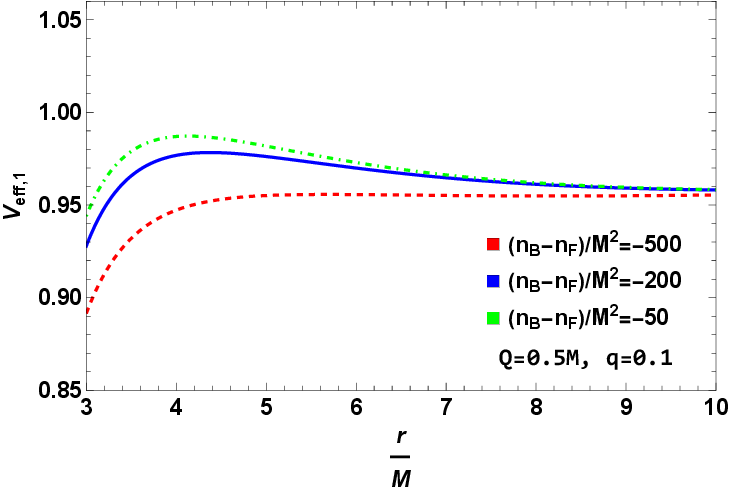}
    \includegraphics[width=0.47\textwidth]{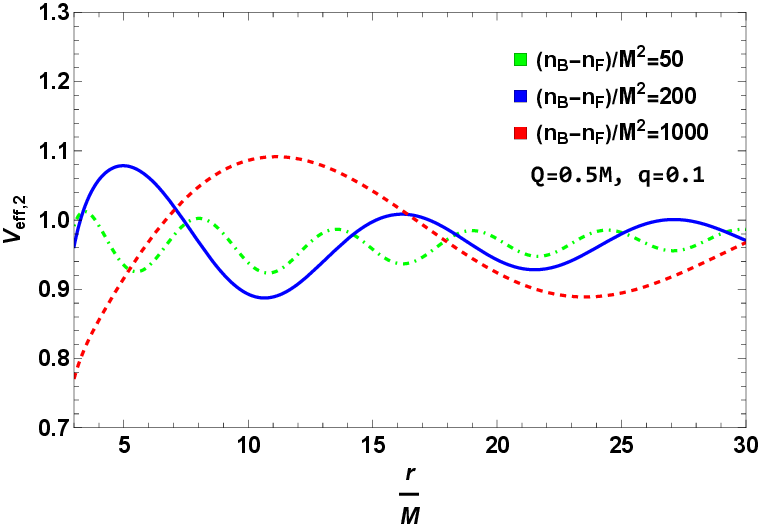}
    \caption{The variation of the effective potential $V_{eff}$ of the charged particles with respect to distance for $\mathcal{L}=4 M$ and for selected negative values of $n_B - n_F$ ($\gamma >0$) (left panel) and positive values of $n_B - n_F$ ($\gamma < 0$) (right panel).}
    \label{Veff-charged-1}
\end{figure}

Fig. \ref{Veff-charged-1} visualizes how the external electric interaction reshapes the effective potential of charged test particles in the Symmergent background. In our test field setup, the electromagnetic contribution enters kinematically through the shift (\ref{energy-charged}) in the radial equation and through the linear term $-q A_t(r)$ in the positive energy branch $V_{\rm eff}^{+}(r)$. The remaining deformation is purely geometric and inherited from the symmergent conformal sector via $A(r), B(r)$ and $C(r)$, which are controlled by $\varphi(r)$, and hence by the sign of $\gamma$. For the negative values of $n_B - n_F$ ($\gamma >0$) (left panel), the Yukawa-suppressed behavior of $\varphi(r)$ leads to a smooth radial profile at larger radii, with the dominant deformation confined to the strong field region. For the positive value of $n_B - n_F > 0$ ($\gamma < 0$)(right panel), the oscillatory behaviour of $\varphi(r)$ induces oscillatory modulations of $V_{eff}$, which persist to larger $r/M$.

\begin{figure}[h!]
   \centering
    \includegraphics[width=0.47\textwidth]{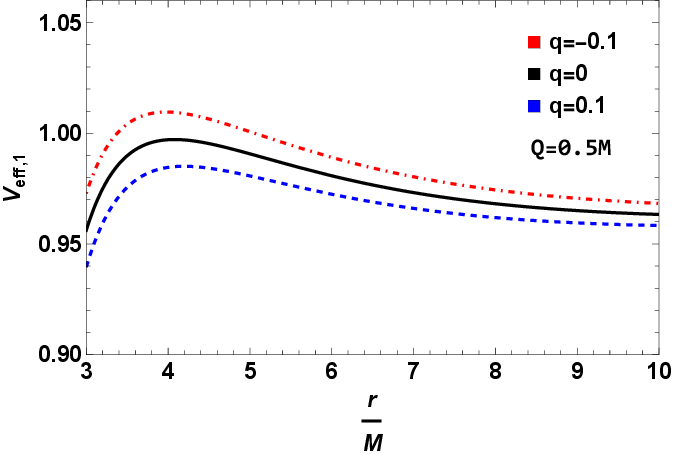}
    \includegraphics[width=0.47\textwidth]{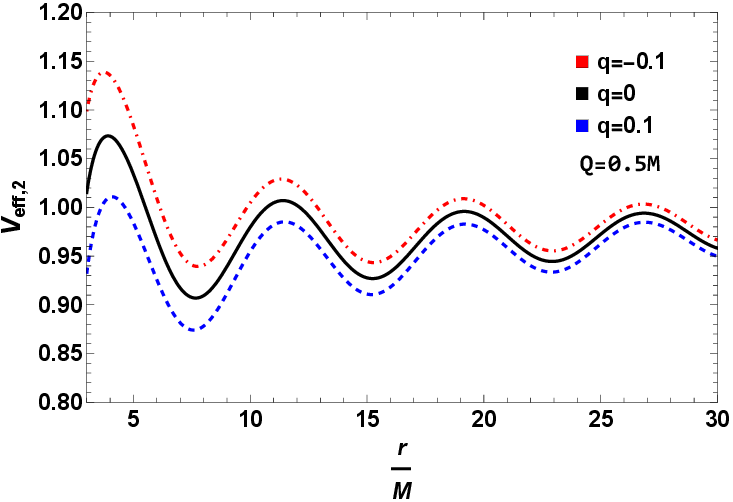}
    \caption{The variation of the effective potential $V_{eff}$ of the charged particles with respect to distance for $Q=0.5 M$, $n_B - n_F = -100$ ($\gamma >0$) (left panel) and $n_B - n_F = 100$ ($\gamma < 0$) (right panel) and for various values of $q$.}
    \label{Veff-charged-2}
\end{figure}

Fig. \ref{Veff-charged-2} isolates the role of the particle’s charge-to-mass ratio q for a fixed external field strength $Q$. It is obvious that from (\ref{eff-pot-charged-corrected}), changing $q$ produces a systematic deformation of the potential landscape through the linear term proportional to $qA_t(r)$, while the background-dependent square root term retains the Symmergent imprint through $A(r)$ and $C(r)$. For the negative values of $n_B - n_F$ ($\gamma >0$), the resulting shift remains smooth and predominantly affects the inner region, whereas for the positive values of $n_B - n_F$ ($\gamma < 0$) the same charge dependence is superimposed on the oscillatory modulation inherited from the conformal sector. Consequently, the charged dynamics exhibits a clear separation of effects: $q$ controls the “Coulomb tilt” of the potential, while $\gamma$ controls whether the geometric correction is short-ranged or oscillatory.

\begin{figure}[h!]
   \centering
    \includegraphics[width=0.475\textwidth]{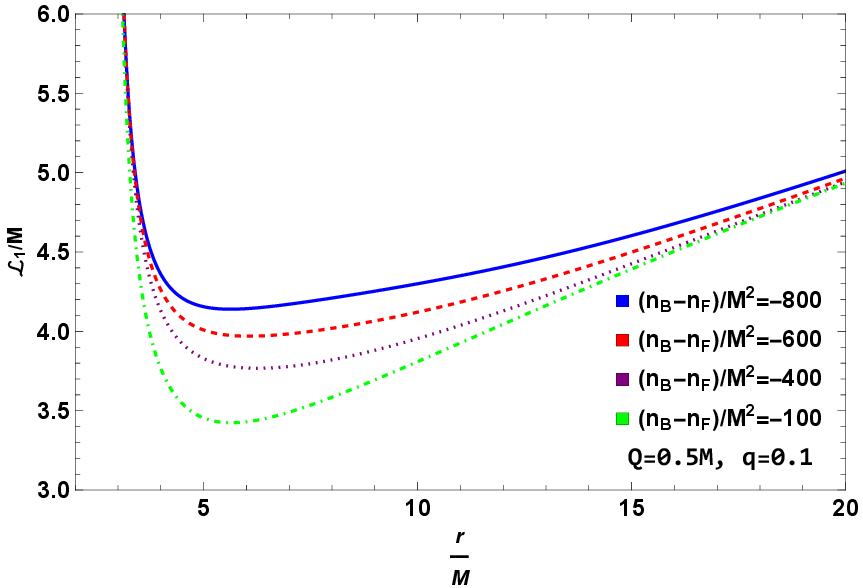}
      \includegraphics[width=0.475\textwidth]{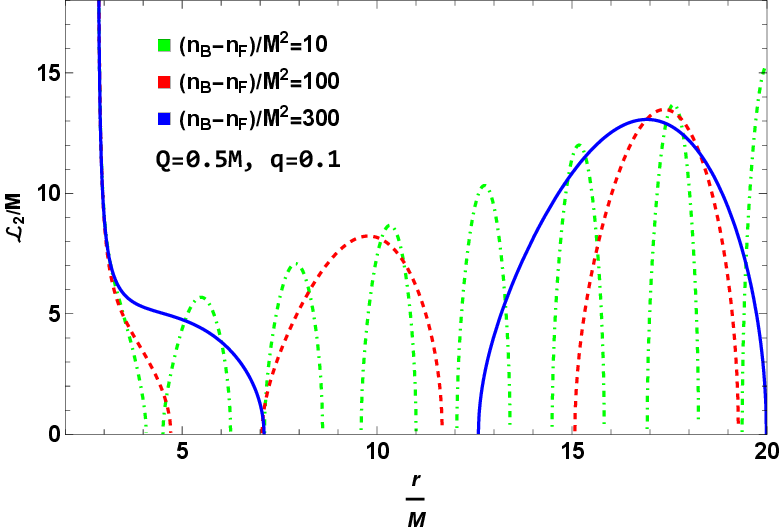}
    \caption{The variation of the angular momentum of the charged particles $\mathcal{L}$ with respect to distance for different values of $n_B - n_F < 0$ ($\gamma >0$) (left panel) and positive value of $n_B - n_F > 0$ ($\gamma < 0$) (right panel).}
    \label{Ang-mom-charged}
\end{figure}

Fig. \ref{Ang-mom-charged}  shows the angular momentum requirement for circular motion of charged particles. Relative to the neutral case, the presence of the external electric field modifies the circular orbit condition, leading to charge-dependent shifts in $\mathcal{L}(r)$. In the $\gamma >0$ branch, the profile remains comparatively smooth at larger radii, consistent with Yukawa suppression of $\varphi(r)$. In the $\gamma < 0$ branch, the oscillatory conformal tail induces oscillatory features in $\mathcal{L}(r)$, implying that the existence of circular orbits can vary sensitively with radius and parameters. These trends are the charged particle counterpart of the neutral orbit behavior, with an additional deformation governed by $q A_t(r)$.

The ISCO is then obtained by imposing, in addition,
\begin{align}
\partial_{rr}V_{\rm eff}^{+}=0 .
\end{align}
For the illustrative numerical analysis we use the Coulomb-type test potential
\begin{align}
A_t(r)=\frac{Q}{r},
\qquad
A_t'(r)=-\frac{Q}{r^2}.
\end{align}
Since the background metric itself is not electrically charged, $Q$ should be
interpreted as the strength of an external test electric field rather than as a
black-hole charge.

\begin{figure}[h!]
   \centering
    \includegraphics[width=0.473\textwidth]{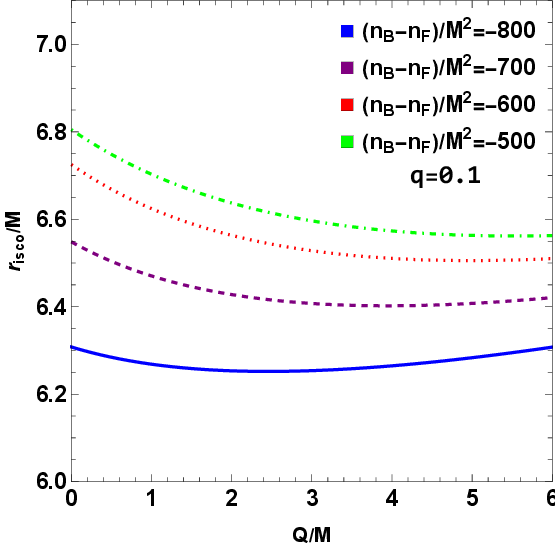}
    \includegraphics[width=0.48\textwidth]{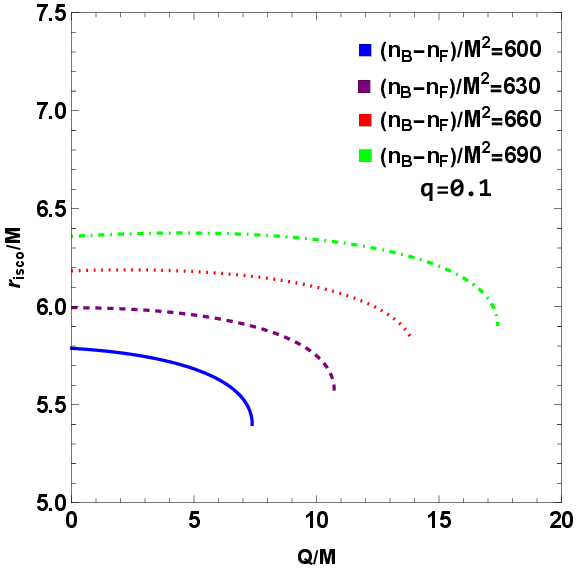}
    \caption{The variation of the ISCO radius $r_{\rm ISCO}$ with charge for $q=0.1$ and for negative values of  $n_B - n_F$ ($\gamma >0$) (left panel) and positive values of $n_B - n_F > 0$ ($\gamma < 0$) (right panel).}
    \label{rISCO-charged-1}
\end{figure}

Fig. \ref{rISCO-charged-1} quantifies how the innermost stable circular orbit responds to the strength $Q$ of the external Coulomb-type test field at fixed charge-to-mass ratio $q$. The ISCO is determined by the stability condition imposed on top of the circularity constraints, so it is particularly sensitive to the detailed shape of $V_{eff}^+(r)$, including both the Coulomb term $q A_t(r)$ and the Symmergent deformation of the metric potentials. In the $\gamma >0$ branch, the dependence of $r_{\rm ISCO}$ on $Q/M$ is smooth and monotonic over the shown parameter range, reflecting the short-ranged nature of $\varphi(r)$. In the $\gamma <0$ branch, the oscillatory conformal sector can generate a more intricate parameter dependence, including non-monotonic features, since the stability conditions probe derivatives of the effective potential and thus amplify oscillatory modulations. This demonstrates that external electromagnetic interactions and the Symmergent conformal sector can interplay nontrivially in setting the innermost stable orbital scale.

\begin{figure}[h!]
   \centering
    \includegraphics[width=0.465\textwidth]{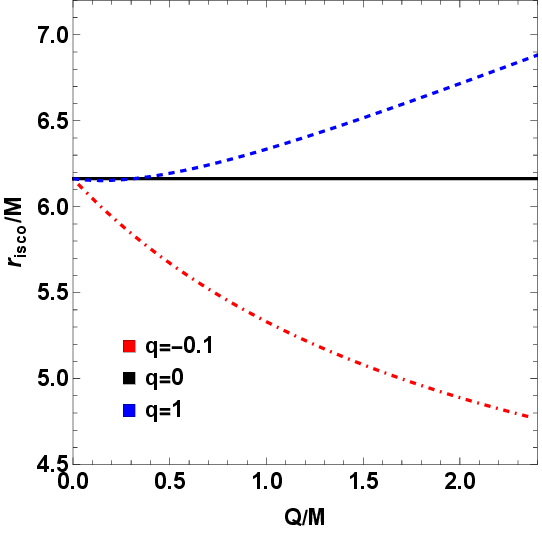}
    \includegraphics[width=0.48\textwidth]{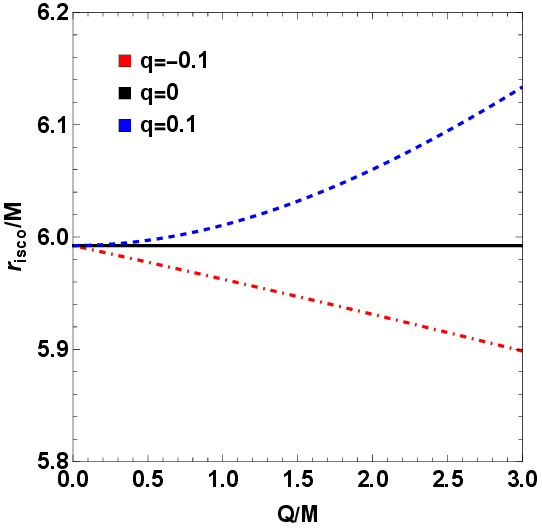}
    \caption{The variation of $r_{\rm ISCO}$ with charge for different values of $q$ for $n_B - n_F = -100$ ($\gamma >0$) (left panel) and positive value of $n_B - n_F = 600$ ($\gamma < 0$) (right panel).}
    \label{rISCO-charged-2}
\end{figure}

Fig. \ref{rISCO-charged-2} highlights the role of the charge sign by comparing $r_{\rm ISCO}$ for positive, vanishing, and negative values of $q$. In both branches, the neutral particle limit is recovered at $Q=0$, where the Coulomb coupling $qA_t(r)=q Q/r$ vanishes and all charged particle curves meet the corresponding neutral symmergent ISCO value. For $q=0$, the ISCO radius remains independent of $Q$, as expected, since a neutral test particle does not couple to the external electric field. In the $\gamma >0$ branch, the effect of the Coulomb coupling is pronounced. Positive and negative values of $q$ shift the ISCO radius in opposite directions: one branch pushes the marginally stable orbit outward, while the other pulls it inward. This reflects the fact that the charged particle effective potential depends on the product $qQ$, so changing the sign of $q$ reverses the electromagnetic contribution to the radial force balance. The smooth behavior of the curves is consistent with the Yukawa-suppressed conformal deformation in this branch. In the $\gamma <0$ branch, the same qualitative charge splitting persists, but the variation of $r_{\rm ISCO}$ with $Q/M$ is milder over the displayed range. The oscillatory conformal sector modifies the stability condition, while the Coulomb interaction separates the positive and negative $q$ trajectories around the $q=0$ curve. Thus, the figure demonstrates that the charged particle ISCO is controlled by the combined effect of the symmergent geometry and the electromagnetic coupling, with the sign of $qQ$ determining whether the ISCO is shifted inward or outward.

\begin{figure}[h!]
   \centering
    \includegraphics[width=0.47\textwidth]{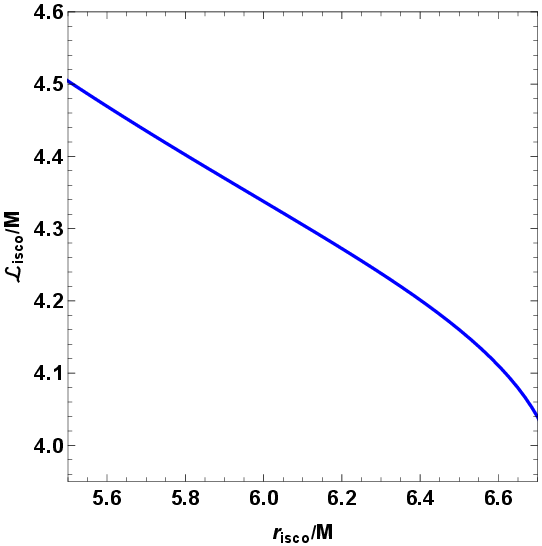}
    \includegraphics[width=0.465\textwidth]{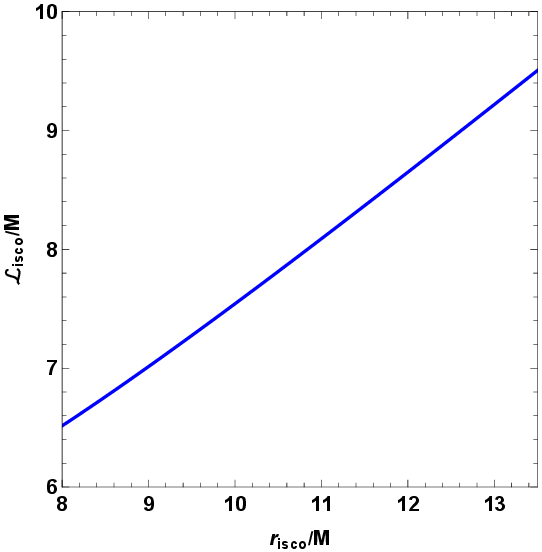}
    \caption{The relation between the ISCO radius and the angular momentum at the ISCO for charged particles for negative values of $n_B - n_F < 0$ ($\gamma >0$) (left panel) and positive values of $n_B - n_F > 0$ ($\gamma < 0$) (right panel).}
    \label{Ang-mom-rISCO-charged}
\end{figure}

Fig. \ref{Ang-mom-rISCO-charged} shows the relation between the ISCO radius and the angular momentum evaluated at the ISCO for charged test particles. The ISCO radius is obtained from the stability condition, while $\mathcal L_{\rm isco}$ is computed by substituting this radius into the charged circular orbit angular momentum expression. Therefore, the figure provides a direct measure of how the angular momentum threshold for stable charged circular motion changes as the ISCO location is shifted. In the $n_B-n_F < 0$ branch $(\gamma>0)$, $\mathcal L_{\rm isco}/M$ decreases as $r_{\rm isco}/M$ increases. This indicates that, in this branch, moving the charged particle ISCO outward does not require a larger angular momentum. Instead, the smooth symmergent deformation together with the Coulomb contribution $qA_t(r) = qQ_e/r$ lowers the angular momentum threshold along the displayed branch. In the $n_B-n_F > 0$ branch $(\gamma<0)$, the trend is reversed: $\mathcal L_{\rm isco}/M$ increases monotonically with $r_{\rm isco}/M$. Thus, larger ISCO radii require larger angular momentum to sustain stable charged circular motion. This behavior reflects the stronger influence of the oscillatory conformal sector on the charged particle stability condition. The opposite slopes in the two panels show that the sign of the Bose-Fermi imbalance qualitatively changes the correlation between the ISCO location and the angular momentum scale.

\begin{figure}[h!]
   \centering
    \includegraphics[width=0.465\textwidth]{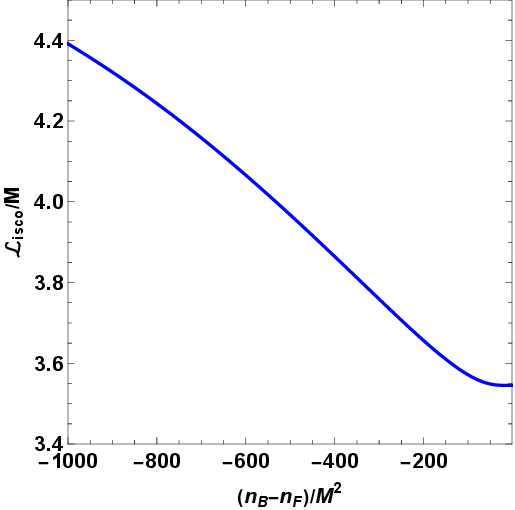}
    \includegraphics[width=0.48\textwidth]{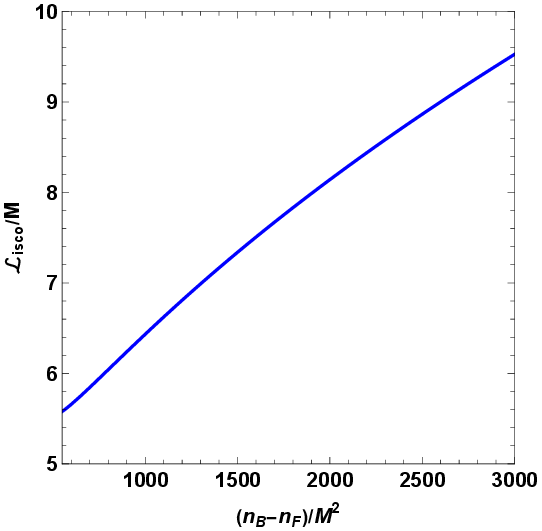}
    \caption{The variation of the angular momentum at ISCO  of charged particles with negative values of $n_B - n_F < 0$ ($\gamma >0$) (left panel) and positive values of $n_B - n_F > 0$ ($\gamma < 0$) (right panel).}
    \label{Ang-mom-ISCO-charged}
\end{figure}

Fig. \ref{Ang-mom-ISCO-charged} shows the variation of the angular momentum at the ISCO for charged test particles as a function of the Bose-Fermi imbalance parameter. Since the charged particle circular orbit condition depends both on the symmergent metric functions and on the Coulomb coupling $qA_t(r) = qQ_e/r$, $\mathcal L_{\rm isco}$ provides a useful diagnostic of how the particle content parameter modifies the angular momentum threshold for stable charged motion. In the $n_B-n_F < 0$ branch $(\gamma>0)$, $\mathcal L_{\rm isco}$ decreases as $(n_B-n_F)/M^2$ approaches zero from below. This indicates that, along this branch, the charged particle requires progressively less angular momentum to remain on the innermost stable circular orbit. The smooth decrease is consistent with the non-oscillatory character of the conformal deformation in the $\gamma>0$ branch. In the $n_B-n_F > 0$ branch $(\gamma<0)$, the behavior is opposite: $\mathcal L_{\rm isco}$ increases monotonically with $n_B-n_F/M^2$. Thus, in the oscillatory conformal branch, increasing the Bose-Fermi imbalance raises the angular momentum threshold required for charged-particle stability. The opposite trends in the two panels show that the sign of $n_B-n_F$ qualitatively changes the charged particle orbital response.

\begin{figure}[h!]
   \centering
   \includegraphics[width=0.475\textwidth]{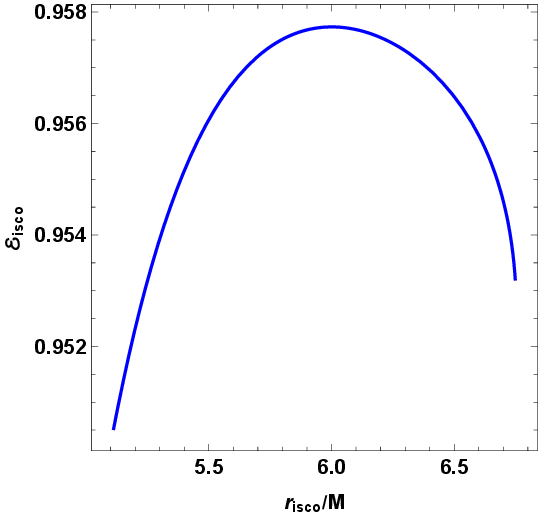}
    \includegraphics[width=0.48\textwidth]{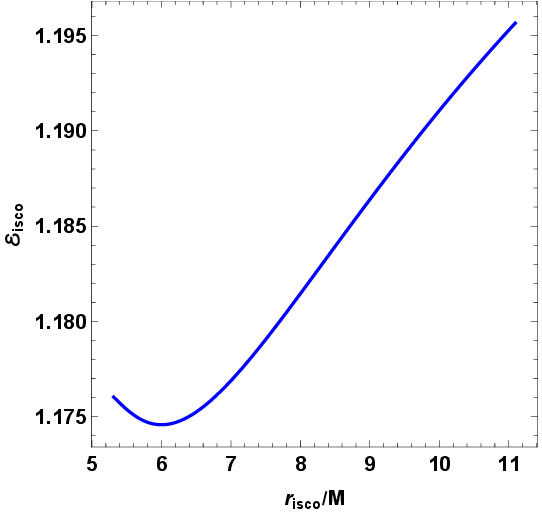}
    \caption{The variation of the energy at the ISCO of charged particles with respect to $r_{\rm ISCO}$ for negative values of $n_B - n_F < 0$ ($\gamma >0$) (left panel) and positive values of $n_B - n_F > 0$ ($\gamma < 0$) (right panel).}
    \label{energy-ISCO-charged}
\end{figure}

Fig. \ref{energy-ISCO-charged} shows the conserved specific energy evaluated at the ISCO as a function of $r_{\rm isco}$ for charged test particles. The energy is obtained from the positive effective-potential branch after substituting the corresponding ISCO angular momentum and ISCO radius. In the $n_B-n_F < 0$ branch $(\gamma>0)$, $\mathcal E_{\rm isco}$ remains below unity and exhibits a non-monotonic profile. This indicates that the corresponding stable charged orbits are bound, while the peak reflects the fact that the ISCO energy is controlled by the combined effect of the symmergent metric functions, the angular momentum threshold, and the Coulomb term $qA_t(r)=qQ_e/r$, rather than by the orbital radius alone.  In the $n_B-n_F > 0$ branch $(\gamma<0)$, $\mathcal E_{\rm isco}$ is larger than unity over the displayed range and increases with $r_{\rm isco}$ after a shallow minimum. This behavior indicates that the oscillatory conformal sector can significantly raise the energy required for marginally stable charged circular motion. Unlike the $\gamma>0$ branch, where the charged ISCO configurations remain bound, the $\gamma<0$ branch corresponds to energetically more costly or unbound configurations in the chosen parameter range. The contrast between the two panels shows that the sign of the Bose-Fermi imbalance qualitatively changes the energetic cost of charged-particle orbital stability.

\begin{figure}[h!]
   \centering
   \includegraphics[width=0.475\textwidth]{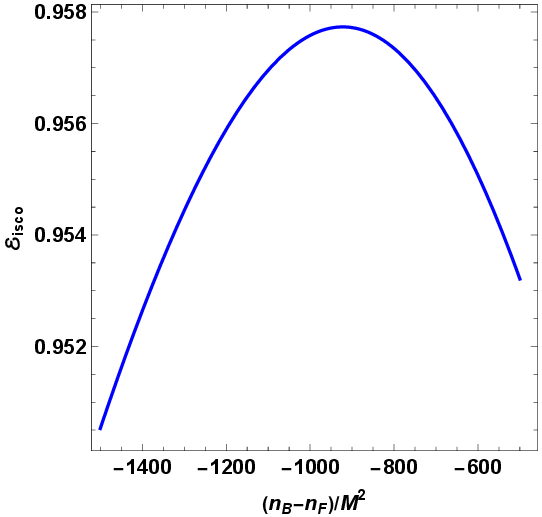}
    \includegraphics[width=0.48\textwidth]{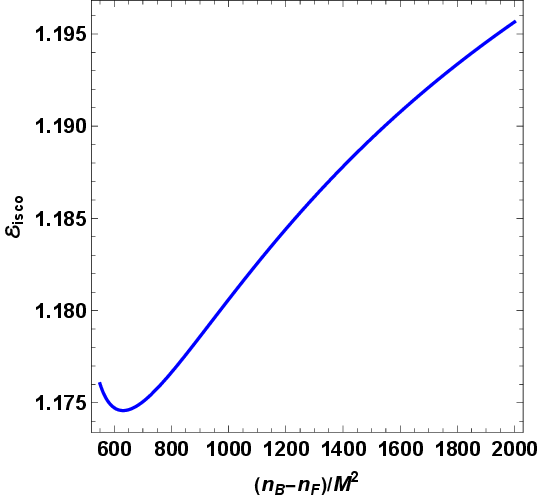}
    \caption{The variation of the energy at the ISCO of charged particles with negative values of $n_B - n_F < 0$ ($\gamma >0$) (left panel) and positive values of $n_B - n_F > 0$ ($\gamma < 0$) (right panel).}
    \label{energy-ISCO-nM-charged}
\end{figure}

Fig. \ref{energy-ISCO-nM-charged} shows the variation of the conserved specific energy at the ISCO, $\mathcal E_{\rm isco}$, with respect to the Bose-Fermi imbalance parameter for charged test particles. The energy is evaluated from the positive effective-potential branch after substituting the corresponding ISCO radius and angular momentum. Thus, $\mathcal E_{\rm isco}$ encodes both the symmergent modification of the background geometry and the electromagnetic contribution through the Coulomb coupling $q A_t(r) = q Q_e/r$. For $n_B - n_F < 0$, $\mathcal E_{\rm isco}$ remains below unity and displays a smooth non-monotonic behavior. This indicates that the corresponding charged particle ISCO configurations are bound, while the peak reflects the parameter range in which the combined geometric and Coulomb contributions maximize the energy required for stable circular motion. For $n_B - n_F > 0$, is larger than unity over the displayed range and increases after a shallow minimum. This shows that the oscillatory conformal branch can substantially raise the energetic cost of charged particle stability. The contrast between the two panels demonstrates that the sign of the Bose--Fermi imbalance qualitatively affects not only the ISCO radius and angular momentum, but also the conserved energy associated with the innermost stable charged-particle orbit.

\begin{figure}[h!]
   \centering
   \includegraphics[width=0.48\textwidth]{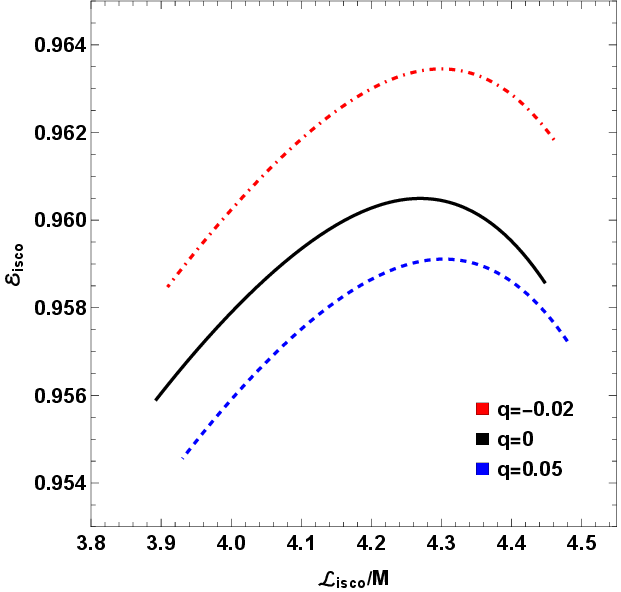}
    \includegraphics[width=0.48\textwidth]{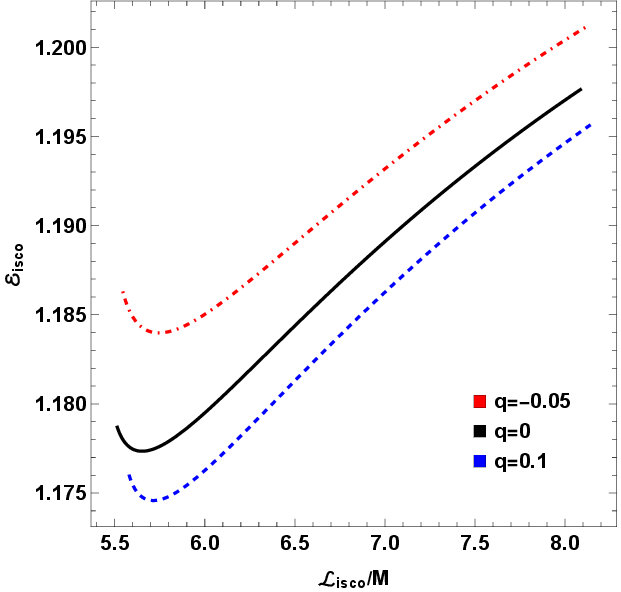}
    \caption{The relation between the specific energy $\mathcal E_{\rm isco}$ and the angular momentum $\mathcal L_{\rm isco}$ at the ISCO for charged particles for negative values of $n_B - n_F < 0$ ($\gamma >0$) (left panel) and positive values of $n_B - n_F > 0$ ($\gamma < 0$) (right panel).}
    \label{E-L-ISCO-charged}
\end{figure}

Fig. \ref{E-L-ISCO-charged} shows the relation between the specific energy and the angular momentum evaluated at the ISCO for charged test particles. The different curves correspond to different values of the charge-to-mass ratio $q$, while the external electric field parameter is kept fixed. In the $n_B - n_F < 0$ branch, $\mathcal E_{\rm isco}$ remains below unity for all displayed values of $q$, indicating bound stable charged particle configurations. The curves exhibit a smooth non-monotonic profile: the energy first increases with $\mathcal L_{\rm isco}$, reaches a maximum, and then decreases. This shows that, in this branch, increasing the angular momentum threshold does not monotonically increase the ISCO energy. Instead, the energy is determined by the combined contribution of the symmergent deformation, the Coulomb term $q A_t(r) = q Q_e/r$, and the ISCO radius selected by the stability condition. In the $n_B - n_F > 0$ branch, the corresponding energies are larger than unity over the displayed range. This indicates that the stable charged configurations in this branch are energetically unbound, or at least require more energy than the rest-mass threshold. After a shallow minimum, $\mathcal E_{\rm isco}$ increases with $\mathcal L_{\rm isco}$, showing that higher angular momentum thresholds are associated with energetically more costly marginally stable orbits. The separation between the curves demonstrates the sensitivity of the ISCO energy-angular-momentum relation to the sign and magnitude of $q$. Thus, the Coulomb coupling shifts the energetic cost of maintaining marginal stability, while the sign of the Bose-Fermi imbalance determines whether the branch supports bound or unbound ISCO configurations.

\subsection{Dynamics of Spinning Particles}
\subsubsection{Equations of motion of spinning particles}

In this section, we briefly review the theoretical framework for the motion of spinning test particles in the gravitational field of compact objects. While a spinless particle follows the usual geodesic equations, the presence of intrinsic spin leads to non-geodesic dynamics due to spin–curvature coupling. The motion of a massive spinning particle is therefore governed by the Mathisson-Papapetrou-Dixon (MPD) equations, which can be written as~\cite{MPD-eq1,MPD-eq2}:
\begin{align}
\label{MPD-eq1}
\frac{Dp^\alpha}{d\lambda}&=-\frac{1}{2}R^\alpha_{\;\;\beta\delta\sigma}u^\beta S^{\delta\sigma},\\
\frac{D S^{\alpha\beta}}{d\lambda}&=p^\alpha u^\beta-p^\beta     u^\alpha
\label{MPD-eq2}
\end{align}
where $D/d\lambda \equiv u^\alpha \nabla_\alpha$ denotes the covariant derivative projected along the particle's worldline, with $u^\mu = dx^\mu/d\lambda$ the corresponding four-velocity. Here $p^\alpha$ is the particle's canonical four-momentum, $R^\alpha{}_{\beta\delta\sigma}$ is the Riemann curvature tensor, and $\lambda$ is an affine parameter. The antisymmetric spin tensor $S^{\alpha\beta}$ satisfies $S^{\alpha\beta} = -S^{\beta\alpha}$.
Clearly, Eq.~(\ref{MPD-eq1},\ref{MPD-eq2}) reduces to the standard geodesic equation of general relativity when the spin tensor components $S^{\alpha\beta}$ vanish. In that limit, the corresponding differential equation takes the form
 \begin{align}
\frac{Dp^\alpha}{d\lambda}=0 \, .
 \end{align}
To solve the system of equations in Eq.~(\ref{MPD-eq1},\ref{MPD-eq2}), one must impose an additional constraint the so-called spin-supplementary condition (SSC) which specifies the particle's center-of-mass worldline. Several SSC choices have been discussed in the literature (see, e.g.,~\cite{tulczyjew1959motion,1998PhRvD..58f4005S,Pirani:1956tn}). In this work, we adopt the Tulczyjew SSC ~\cite{tulczyjew1959motion}, defined by
\begin{align}
 S^{\alpha\beta}p_\alpha=0.
\end{align}
The MPD equations supplemented by the SSC imply two independent conserved quantities, namely the magnitude of the canonical momentum and the spin of the particle, which can be expressed as
\begin{align}
\label{mom-spin1}
S^{\alpha\beta}S_{\alpha\beta}&=2S^2=2m^2s^2, \\
p^\alpha p_\alpha&=-m^2.
\label{mom-spin2}
\end{align}
In addition to the conserved spin and canonical momentum in Eq.~(\ref{mom-spin1},\ref{mom-spin2}), the motion admits further constants of motion dictated by the spacetime symmetries. For an axisymmetric geometry, two Killing vector fields are present: the timelike Killing vector $\xi^\alpha$ associated with time-translation invariance, and the axial Killing vector $\psi^\alpha$ generating rotations in the azimuthal direction $\phi$. The corresponding conserved quantities can be obtained from
\begin{align}
 p^\alpha \kappa_\alpha-\frac{1}{2}S^{\alpha\beta}\nabla_\beta\kappa_\alpha=p^\alpha \kappa_\alpha-\frac{1}{2}S^{\alpha\beta}\partial_\beta\kappa_\alpha=\text{constant},
\end{align}
where $k^\alpha$ denotes either of the two Killing vector fields, $\xi^\alpha$ or $\psi^\alpha$.

We now turn to the motion of a spinning test particle in the background of an asymptotically-flat symmergent black hole. Our strategy is to formulate an effective potential for the dynamics and use it to characterize circular trajectories.

\subsubsection{Effective potential for spinning particles' motion}

Our goal is to construct the effective potential governing a spinning test particle in the spacetime of asymptotically flat symmergent black-hole exterior. For simplicity, we restrict the motion to the equatorial plane, $\theta=\pi/2$. In a static, spherically symmetric background, the dynamics admits two conserved quantities: the energy $E$ and the total angular momentum $J$ are defined as
\begin{align}
 -E&=p_t-\frac{1}{2}g_{tt,r} S^{tr} \, ,
        \\
        J&=p_\phi-\frac{1}{2}g_{\phi\phi,r} S^{\phi r},
\end{align}
where $J=L+S$ is the total angular momentum, with $S=sm$ and $L=\mathcal{L} m$. 
For equatorial motion with the spin vector orthogonal to the orbital plane, the Tulczyjew condition gives the independent spin-tensor components
\begin{align}
S^{tr}
&=
-\frac{s\,p_\phi}
{\sqrt{-g_{tt}g_{rr}g_{\phi\phi}}},
\label{spin-tensor1}
\\
S^{t\phi}
&=
\frac{s\,p_r}
{\sqrt{-g_{tt}g_{rr}g_{\phi\phi}}},
\label{spin-tensor-middle}
\\
S^{\phi r}
&=
\frac{s\,p_t}
{\sqrt{-g_{tt}g_{rr}g_{\phi\phi}}}.
\label{spin-tensor2}
\end{align}
The component $S^{t\phi}$ vanishes on an exactly circular orbit, for which $p_r=0$, but it is nonzero for generic radial motion and should not be omitted when discussing the full effective radial dynamics.

On a circular orbit, where $p_r=0$ and hence $S^{t\phi}=0$, using Eqs.~\eqref{spin-tensor1} and \eqref{spin-tensor2}, the conserved energy and
total angular momentum can be written as
\begin{align}
-E
&=
p_t
-
\frac{sA'(r)}
{2\sqrt{A(r)C(r)/B(r)}}p_\phi,
\\
J
&=
p_\phi
-
\frac{sC'(r)}
{2\sqrt{A(r)C(r)/B(r)}}p_t .
\end{align}
Solving these two algebraic equations gives
\begin{align}
\label{mom-sol1-corrected}
p_t
&=
\frac{
-4E A(r)C(r)
+
2JsA'(r)\sqrt{A(r)B(r)C(r)}
}
{
4A(r)C(r)
-
s^2B(r)A'(r)C'(r)
},
\\
\label{mom-sol2-corrected}
p_\phi
&=
\frac{
4J A(r)C(r)
-
2EsC'(r)\sqrt{A(r)B(r)C(r)}
}
{
4A(r)C(r)
-
s^2B(r)A'(r)C'(r)
}.
\end{align}
Substituting \eqref{mom-sol1-corrected} and \eqref{mom-sol2-corrected} into
\begin{align}
(p^r)^2
=
-g^{rr}
\left(
g^{tt}p_t^2
+
g^{\phi\phi}p_\phi^2
+
m^2
\right),
\end{align}
one obtains

\begin{align}
\label{mom-parametric-corrected}
\left(\frac{p^r}{m}\right)^2
=
\frac{
\alpha_s\mathcal E^2
+
\delta_s\mathcal E
+
\Gamma_s
}
{\rho_s},
\end{align}
where $\mathcal E=E/m$ and $\mathcal J=J/m$. We emphasize that $p^\mu/m$ is not, in general, identical to the tangent four-velocity $u^\mu$ under the Tulczyjew spin-supplementary condition. The difference begins at quadratic order in spin. Consequently, Eq.~\eqref{mom-parametric-corrected} is a radial momentum equation, not directly an equation for $(u^r)^2$. The timelike character and future-directed orientation of the physical tangent vector must be checked separately using the MPD momentum--velocity relation.

The coefficient functions are
\begin{align}
\alpha_s
&=
4A(r)B(r)
\left[
4C(r)^2
-
s^2B(r)C'(r)^2
\right],
\\
\delta_s
&=
16\mathcal J s B(r)
\sqrt{A(r)B(r)C(r)}
\left[
A(r)C'(r)-C(r)A'(r)
\right],
\\
\rho_s
&=
\left[
4A(r)C(r)
-
s^2B(r)A'(r)C'(r)
\right]^2,
\\
\Gamma_s
&=
4\mathcal J^2B(r)C(r)
\left[
s^2B(r)A'(r)^2
-
4A(r)^2
\right]
-
B(r)\rho_s .
\end{align}
We have denoted the constant term by $\Gamma_s$ to avoid confusion with the
symmergent parameter $\gamma$.

To characterize the radial momentum of a spinning particle, we set $p^r=0$. Equation~\eqref{mom-parametric-corrected} is then quadratic in the specific energy and gives
\begin{align}
V_\pm(r)
=
\frac{
-\delta_s(r)
\pm
\sqrt{\delta_s(r)^2-4\alpha_s(r)\Gamma_s(r)}
}
{2\alpha_s(r)}.
\label{spinning-effective-branches}
\end{align}
For the positive-energy sector, we use
\begin{align}
V_{\rm eff}(r)
\equiv
V_+(r)
=
\frac{
-\delta_s(r)
+
\sqrt{\delta_s(r)^2-4\alpha_s(r)\Gamma_s(r)}
}
{2\alpha_s(r)}.
\label{spinning-effective-positive}
\end{align}
The construction is meaningful only where
\begin{align}
\delta_s^2-4\alpha_s\Gamma_s&\geq0,
\label{spin-discriminant}
\\
4A(r)C(r)-s^2B(r)A'(r)C'(r)&\neq0,
\label{spin-denominator}
\\
\alpha_s(r)&\neq0.
\label{spin-alpha}
\end{align}
The sign of $\alpha_s$ must be checked before translating Eq.~\eqref{mom-parametric-corrected} into inequalities involving $\mathcal E$ and $V_\pm$. Circular momentum configurations satisfy
\begin{align}
p^r&=0,
\label{spinning-circular-conditions}
\end{align}
while marginal stability additionally requires
\begin{align}
\frac{d^2}{dr^2}\left(\frac{p^r}{m}\right)^2=0.
\label{spinning-marginal-condition}
\end{align}
For every candidate orbit, the physical tangent vector obtained from the MPD momentum--velocity relation must remain timelike and future-directed. The pole--dipole approximation further requires the M{\o}ller radius to remain much smaller than the local curvature scale; in the present dimensionless variables this requires $|s|/M\ll1$. Results obtained for $|s|/M$ of order unity should not be interpreted as quantitatively reliable test-particle predictions. Because the effective potential is algebraically cumbersome, its behavior is illustrated graphically only after all of the above admissibility constraints are imposed.

\begin{figure}[h!]
   \centering
    \includegraphics[width=0.46\textwidth]{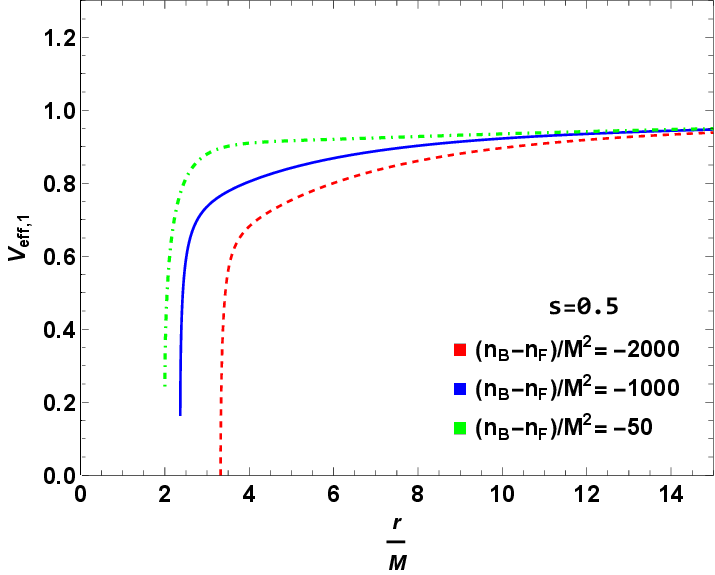}
    \includegraphics[width=0.47\textwidth]{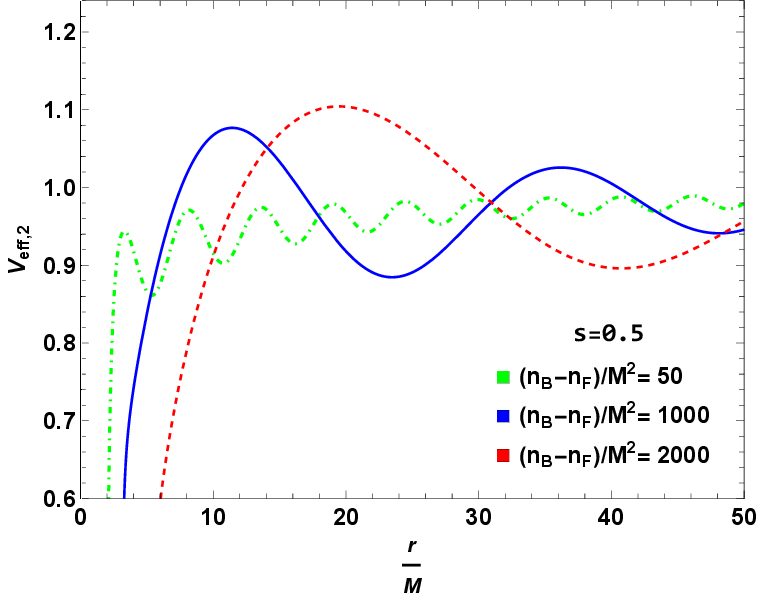}
    \caption{The variation of the effective potential of the spinning particles with respect to distance for different values of $(n_B - n_F)/M^2 < 0$ ($\gamma >0$) (left panel) and positive values of $(n_B - n_F)/M^2 > 0$ ($\gamma < 0$) (right panel).}
    \label{Veff-spin}
\end{figure}

Fig. \ref{Veff-spin} shows how the effective potential governing the radial motion of spinning test particles depends on the underlying Symmergent branch selected by the Bose-Fermi imbalance. In the present setup, spin enters through spin–curvature coupling (MPD dynamics with Tulczyjew SSC), modifying the conserved quantities and the radial momentum relation and hence deforming the effective potential relative to the spinless case. The symmergent correction, on the other hand, is encoded in the conformal deformation $\varphi(r)$, which affects the metric functions $A(r), B(r), C(r)$ and thus the entire potential landscape. For $(n_B-n_F)/M^2 < 0$ ($\gamma > 0$), the Yukawa-suppressed conformal sector leads to a smooth potential profile that rapidly settles as $r/M$ increases, indicating that spin-induced modifications are most relevant in the strong-field region. For $(n_B-n_F)/M^2 > 0$ ($\gamma < 0$), the oscillatory $1/r$ tail of $\varphi(r)$ is imprinted on the effective potential, producing oscillatory modulations whose amplitude and persistence depend on $n_B-n_F$. This is the spinning particle analogue of the neutral/charged cases, with the additional feature that spin–curvature coupling can shift the location and height of the potential barrier that controls turning points and circular orbit structure.

\begin{figure}[h!]
   \centering
    \includegraphics[width=0.465\textwidth]{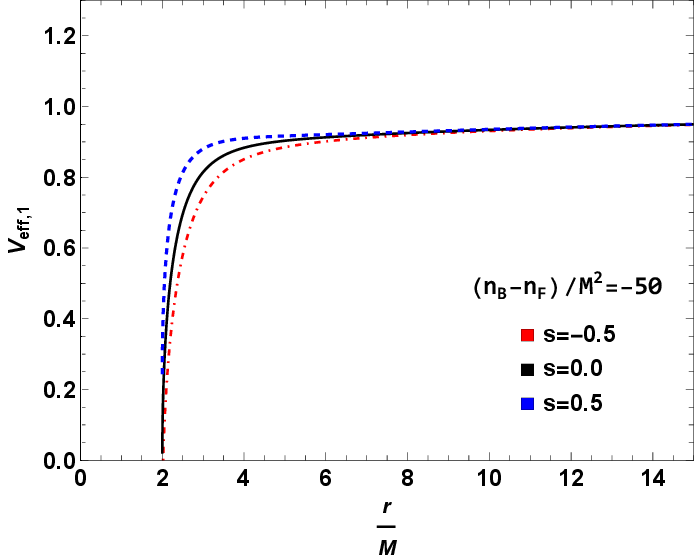}
    \includegraphics[width=0.465\textwidth]{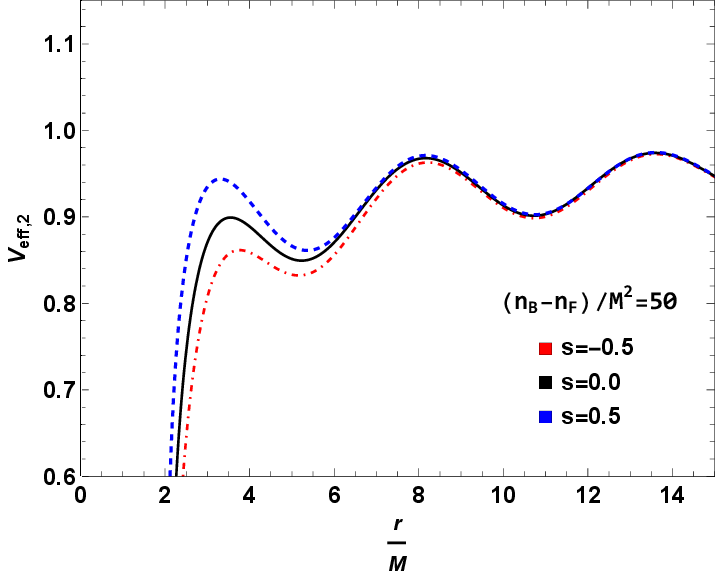}
    \caption{The variation of the effective potential of the spinning particles with respect to distance for different values of spin and for $(n_B - n_F)/M^2 = -50$ ($\gamma >0$) (left panel) and positive values of $(n_B - n_F)/M^2 = 50$ ($\gamma < 0$) (right panel).}
    \label{Veff-spinS}
\end{figure}

Fig. \ref{Veff-spinS} isolates the effect of the spin parameter s on the effective potential at fixed Symmergent branch. Increasing spin changes the strength of spin–curvature coupling and therefore shifts the potential barrier and the turning-point structure. In the $\gamma > 0$ branch, the potential deformation induced by spin remains smooth, and different spin values mainly lead to systematic shifts of the potential profile in the inner region, consistent with the short-ranged nature of $\varphi(r)$. In the $\gamma < 0$ branch, the same spin dependence is superimposed on the oscillatory modulation inherited from the conformal sector. As a result, varying s not only shifts the overall potential but can also change the detailed pattern of local extrema, which in turn affects the existence and stability of circular orbits. This interplay makes the ISCO and the admissible orbital domain particularly sensitive to spin in the oscillatory branch.

\begin{figure}[h!]
   \centering
    \includegraphics[width=0.465\textwidth]{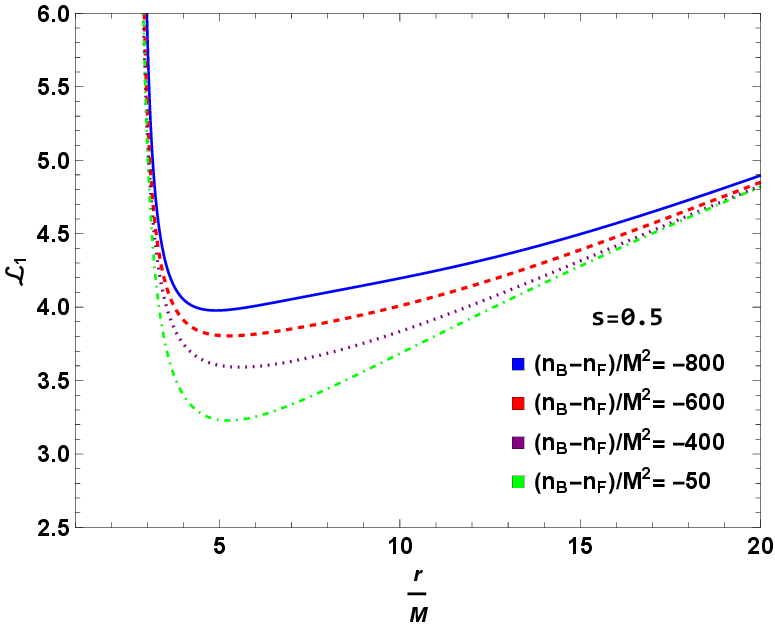}
   \includegraphics[width=0.465\textwidth]{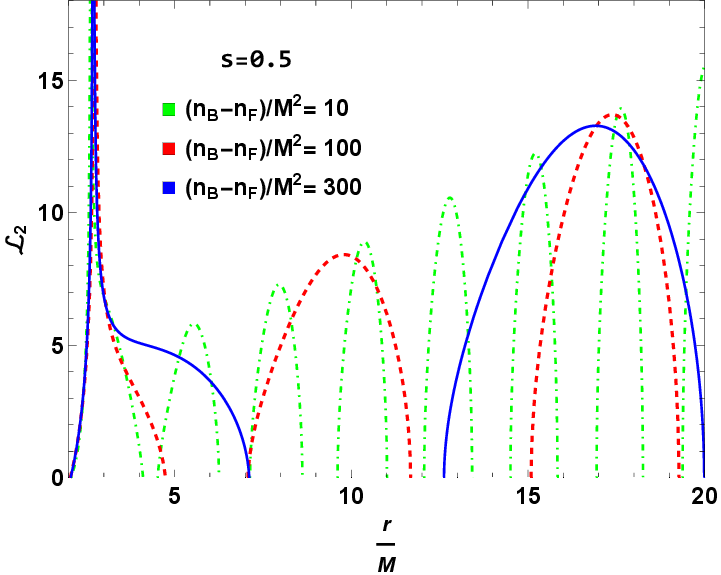}
    \caption{The variation of the angular momentum of the spinning particles $\mathcal{L}$ with respect to distance for fixed spin value $s=0.5$ and for different values of $(n_B - n_F)/M^2 < 0$ ($\gamma >0$)(left panel) and positive value of $(n_B - n_F)/M^2 > 0$ ($\gamma < 0$)(right panel).}
    \label{Ang-mom-spin-1}
\end{figure}

Fig. \ref{Ang-mom-spin-1} shows the angular momentum for circular motion of spinning particles as a function of radius for a fixed spin value. In MPD dynamics, circular trajectories are determined by the simultaneous conditions $p^r=0$ and the extremality of the effective potential, which translate into charge/spin–dependent algebraic relations for the conserved quantities. Relative to the spinless case, spin-curvature coupling shifts the circular-orbit condition and hence modifies $\mathcal{L}(r)$. The Symmergent sector further deforms this relation through $\varphi(r)$. For $\gamma > 0$, the resulting $\mathcal{L}(r)$ curves are smooth and differ mainly in the strong-field region, reflecting the Yukawa suppression of $\varphi(r)$ at large radii. For $\gamma < 0$, oscillations in the conformal sector induce pronounced oscillatory features in $\mathcal{L}(r)$, implying that the existence of circular spinning-particle orbits can vary sensitively with radius and parameters. This behavior anticipates a richer ISCO structure in the oscillatory branch.

\begin{figure}[h!]
   \centering
  \includegraphics[width=0.46\textwidth]{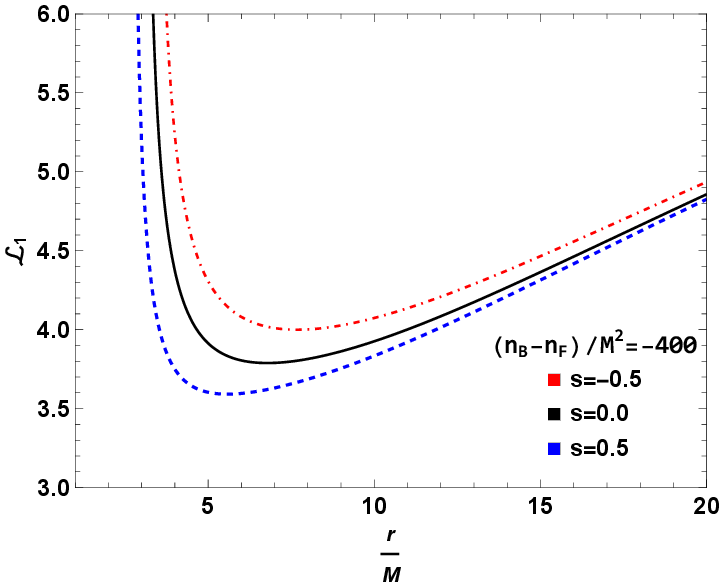}
    \includegraphics[width=0.455\textwidth]{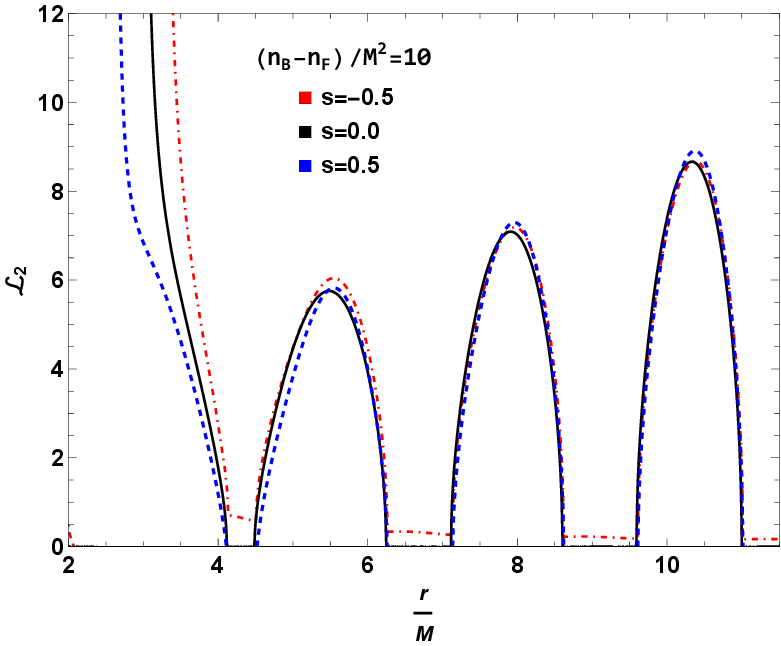}
    \caption{The variation of the angular momentum of the spinning particles with respect to distance for different values of spin and for $(n_B - n_F)/M^2 = -50$ ($\gamma >0$) (left panel) and positive values of $(n_B - n_F)/M^2 = 100$ ($\gamma < 0$) (right panel).}
    \label{Ang-Mom-spinS}
\end{figure}

Fig.~\ref{Ang-Mom-spinS} emphasizes how changing the spin parameter modifies the circular-orbit angular-momentum profile. In the $\gamma>0$ branch, different spin values primarily produce a smooth shift of $\mathcal L(r)$, with the largest effect in the strong-field region because the conformal deformation is Yukawa suppressed at large radius. In the $\gamma<0$ branch, the spin dependence is superimposed on the oscillatory radial modulation inherited from $\varphi(r)$, producing a more structured response. Thus spin--curvature coupling can shift the circular-orbit conditions in both branches, while only the $\gamma<0$ branch carries the long-range oscillatory pattern.

\begin{figure}[htb!]
   \centering
    \includegraphics[width=0.46\textwidth]{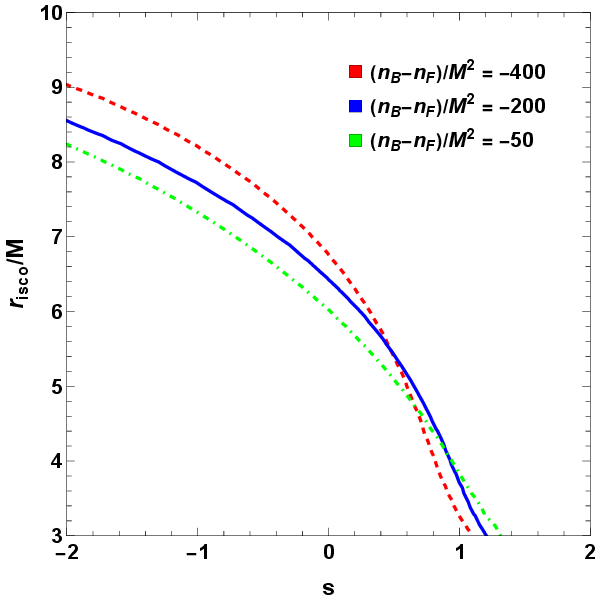}
    \includegraphics[width=0.455\textwidth]{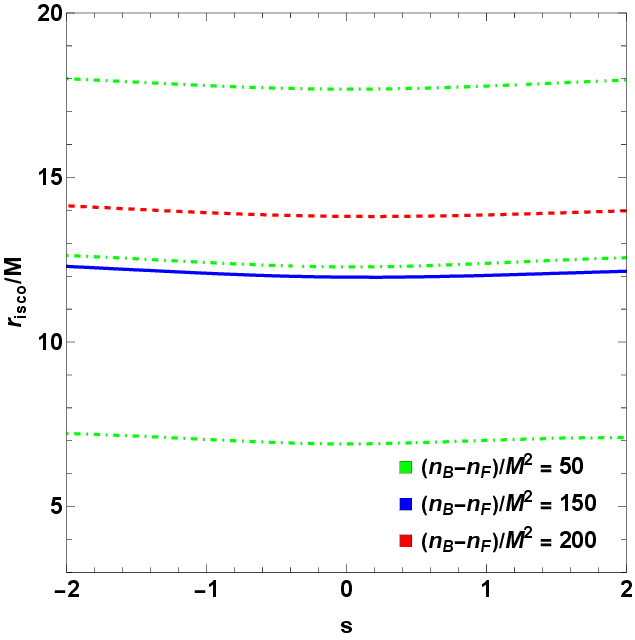}
    \caption{The variation of the ISCO radius of spinning particles with respect to spin for negative values of $(n_B - n_F)/M^2$ ($\gamma >0$)(left panel) and positive values of $(n_B - n_F)/M^2$ ($\gamma < 0$)(right panel).}
    \label{ISCO-radius-spin}
\end{figure}

Fig. \ref{ISCO-radius-spin} shows the ISCO radius as a function of the spin parameter for representative values of $(n_B - n_F)/M^2$. The ISCO is determined by imposing stability in addition to circularity, and therefore it probes derivatives of the effective potential; this makes it highly sensitive to both spin-curvature coupling and to the Symmergent deformation of the metric functions. In the $\gamma > 0$ branch, $r_{\rm ISCO}$ exhibits a clear spin dependence: varying $s$ shifts the stability boundary in a smooth manner, reflecting the fact that the conformal correction is short-ranged and does not introduce long distance structure. In the $\gamma < 0$ branch, the ISCO response can become markedly different: because the background carries oscillatory modulations, the stability conditions may amplify these features, leading to a more intricate dependence of $r_{\rm ISCO}$ on $s$. Overall, the figure demonstrates that spin and the Bose-Fermi-controlled Symmergent branch jointly determine the innermost stable orbital scale.

\begin{figure}[htb!]
   \centering
    \includegraphics[width=0.45\textwidth]{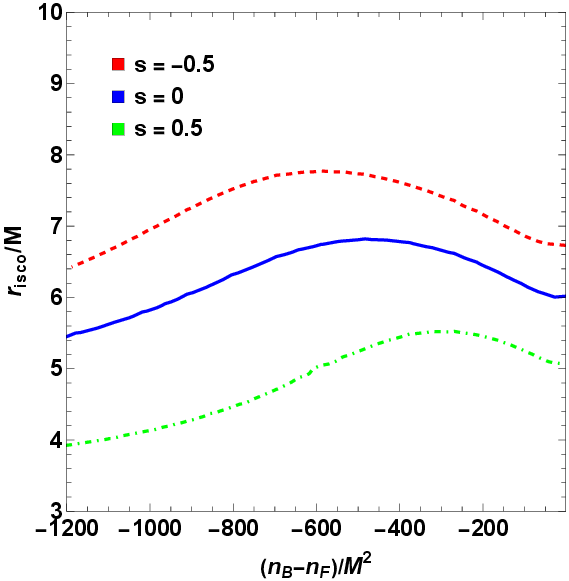}
    \includegraphics[width=0.445\textwidth]{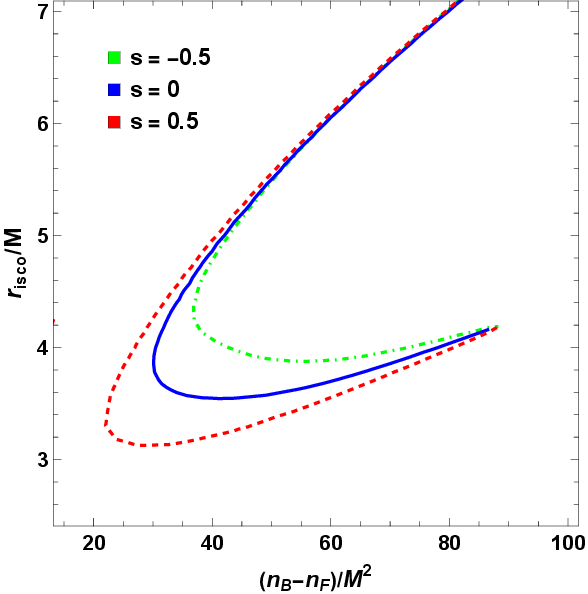}
    \caption{The variation of the ISCO radius of spinning particles with respect to the negative values of $(n_B - n_F)/M^2$ ($\gamma >0$)(left panel) and positive values of $(n_B - n_F)/M^2$ ($\gamma < 0$)(right panel) for various values of spin.}
    \label{ISCO-radius-spin-nBnF}
\end{figure}

Fig. \ref{ISCO-radius-spin-nBnF} directly maps the particle content parameter $n_B - n_F$ to the macroscopic stability boundary $r_{\rm ISCO}$ for several spin values. Scanning $n_B - n_F$ effectively scans the conformal-sector behavior $\varphi(r)$, while changing $s$ modifies the spin-curvature coupling in the MPD equations.  In the $\gamma > 0$ branch, the ISCO varies smoothly with $(n_B - n_F)/M^2$, and different spin values yield correlated shifts consistent with a short-ranged deformation. In the $\gamma < 0$ branch, oscillatory conformal behavior can generate nontrivial structure in the ISCO dependence, and spin can either enhance or partially mitigate these shifts depending on its sign and magnitude. The key physical point is that the Bose–Fermi imbalance is imprinted not only in asymptotic behavior but also in strong-field orbital stability, with spin acting as an additional control parameter.

\begin{figure}[htb!]
   \centering
    \includegraphics[width=0.455\textwidth]{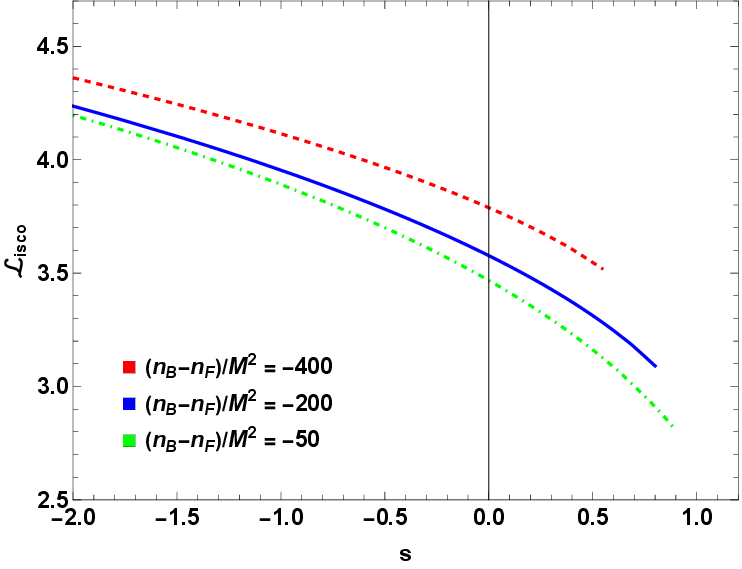}
    \includegraphics[width=0.45\textwidth]{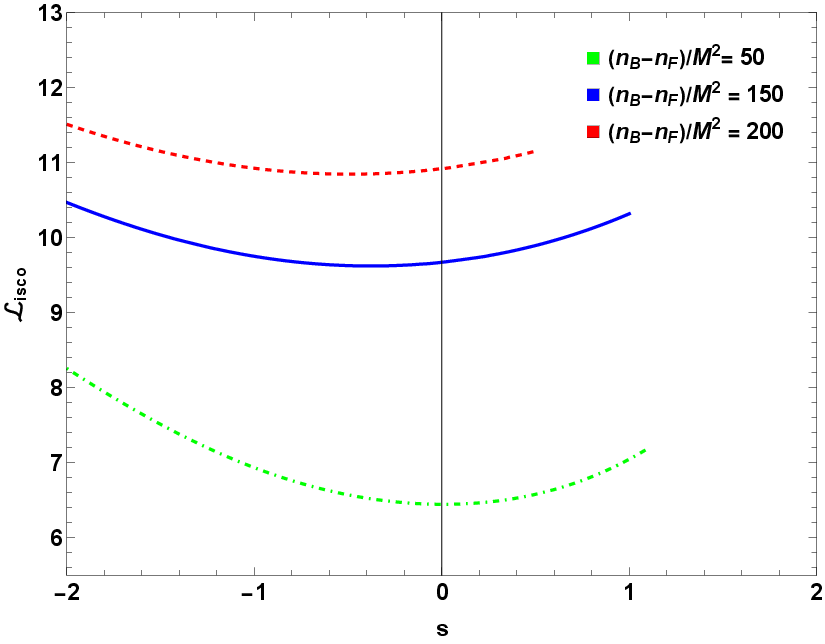}
    \caption{The variation of the angular momentum of spinning particles at ISCO radius  with respect to spin for negative values of $(n_B - n_F)/M^2$ ($\gamma >0$)(left panel) and positive values of $(n_B - n_F)/M^2$ ($\gamma < 0$)(right panel).}
    \label{Ang-mom-ISCO-spin}
\end{figure}

Fig.\ref{Ang-mom-ISCO-spin} shows how the angular momentum at the ISCO depends on the spin parameter in the two Symmergent branches. Because $\mathcal{L}_{isco}$ is obtained by enforcing circularity and stability simultaneously, it captures the combined effect of spin-curvature coupling and the Symmergent deformation of the background geometry. In the $\gamma > 0$ branch, the dependence of $\mathcal{L}_{isco}$ on $s$ is smooth and monotonic over the shown range, consistent with the Yukawa-suppressed nature of the conformal correction. In the $\gamma < 0$ branch, the oscillatory conformal sector can translate into a more structured response, with the ISCO angular momentum threshold exhibiting a nontrivial s-dependence. This is particularly relevant for collision phenomenology, since $\mathcal{L}_{isco}$ sets the angular-momentum scale of the innermost long-lived orbits that can seed high-energy interactions in the strong-field region.

\begin{figure}[htb!]
   \centering
    \includegraphics[width=0.46\textwidth]{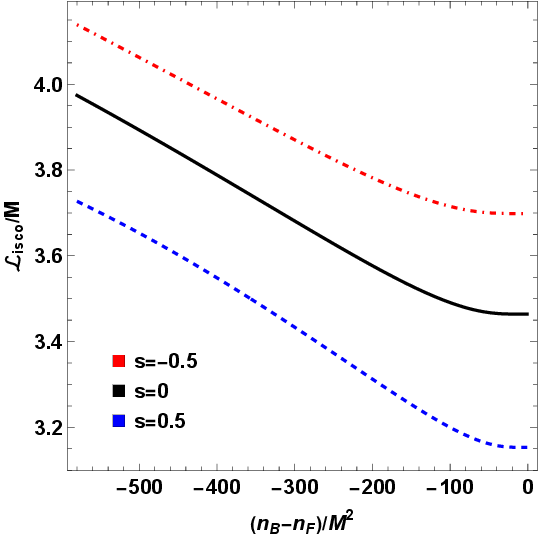}
    \includegraphics[width=0.465\textwidth]{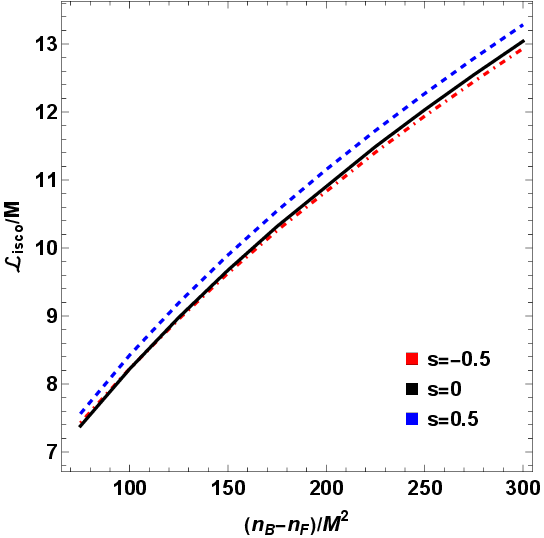}
    \caption{The variation of the angular momentum of spinning particles at the ISCO for negative values of $(n_B - n_F)/M^2$ ($\gamma >0$) (left panel) and positive values of $(n_B - n_F)/M^2$ ($\gamma < 0$) (right panel).}
    \label{Ang-mom-ISCO-spin-s}
\end{figure}

Fig. \ref{Ang-mom-ISCO-spin-s}  shows the angular momentum of spinning particles at the ISCO as a function of the Symmergent parameter $(n_B-n_F)/M^2$. The left panel corresponds to the negative branch, $(n_B-n_F)<0$ or equivalently $\gamma>0$, while the right panel represents the positive branch, $(n_B-n_F)>0$ or $\gamma<0$. For the $\gamma>0$ branch, $\mathcal L_{\rm isco}/M$ decreases as $(n_B-n_F)/M^2$ approaches zero from below. This indicates that larger negative $(n_B-n_F)/M^2$ requires a larger angular momentum for a spinning particle to remain on a stable circular orbit. The spin dependence is clearly ordered: anti-aligned spin, $s=-0.5$, gives the largest ISCO angular momentum, the spinless case lies in between, and aligned spin, $s=0.5$, gives the smallest value. Thus, in this branch, spin-curvature coupling shifts the ISCO angular momentum threshold in a systematic and regular way. The positive branch exhibits a qualitatively different behavior. For $\gamma<0$, $\mathcal L_{\rm isco}/M$ increases monotonically with $(n_B-n_F)/M^2$, showing that the angular momentum required to sustain the stable orbit grows as the positive $(n_B-n_F)/M^2$ becomes larger. Moreover, the spin ordering is reversed with respect to the negative branch: the $s=0.5$ curve lies above the spinless case, whereas $s=-0.5$ gives the smallest angular momentum. This reversal reflects the different way in which spin-curvature coupling combines with the symmergent deformation in the two branches. The comparison between the two panels therefore shows that the sign of $(n_B-n_F)/M^2$ has a direct impact on the ISCO angular momentum scale. The $\gamma>0$ branch produces a smooth, relatively mild deformation of the Schwarzschild-like behavior, whereas the $\gamma<0$ branch leads to a much stronger enhancement of $\mathcal L_{\rm isco}$ over the displayed range. Hence, the ISCO angular momentum of spinning particles provides a sensitive diagnostic of the underlying symmergent branch structure.

\begin{figure}[h!]
   \centering
    \includegraphics[width=0.47\textwidth]{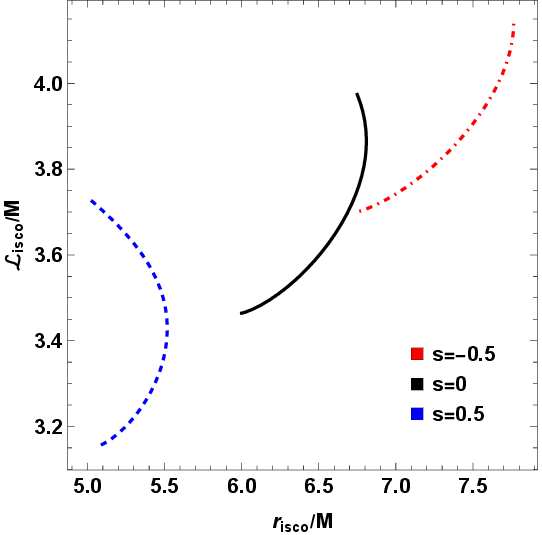}
    \includegraphics[width=0.465\textwidth]{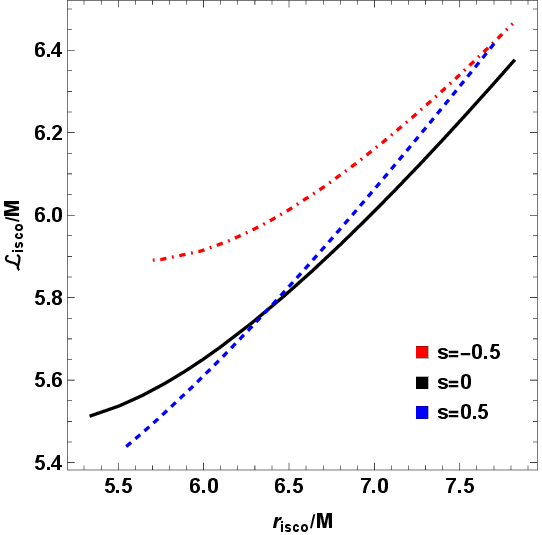}
    \caption{The relation between the ISCO radius and the angular momentum at the ISCO for spinning particles for negative values of $n_B - n_F < 0$ ($\gamma >0$) (left panel) and positive values of $n_B - n_F > 0$ ($\gamma < 0$) (right panel).}
    \label{Ang-mom-rISCO-spin}
\end{figure}

Fig. \ref{Ang-mom-rISCO-spin} shows the relation between the ISCO radius and the angular momentum of spinning particles at the ISCO. In contrast to the previous figure, where the angular momentum was plotted directly as a function of the Symmergent parameter, this figure displays the parametric relation between $r_{isco}$ and $\mathcal{L}_{isco}$. It therefore illustrates how the location of the stable circular orbit and the angular momentum threshold vary together. In the negative branch, $n_B-n_F < 0$ $(\gamma > 0)$, the three spin configurations occupy clearly separated regions of the $(r_{isco}, \mathcal{L}_{isco})$ plane. The $s=−0.5$ curve is shifted toward larger ISCO radii and larger angular momenta, whereas the $s=0.5$ curve lies at smaller radii and lower angular momenta. The spinless case remains between these two behaviors. This shows that, in the $\gamma > 0$ branch, spin–curvature coupling systematically changes both the ISCO location and the angular momentum required to sustain stable circular motion. In the positive branch, $n_B-n_F > 0$ $(\gamma < 0)$, the angular momentum increases with the ISCO radius for all spin configurations. The relation is more monotonic and direct: larger ISCO radii require larger values of $\mathcal{L}_{isco}$. The different spin curves remain distinguishable, although they are closer to one another than in the negative branch. This indicates that the positive branch preserves a strong positive correlation between orbital radius and angular momentum, while spin orientation produces additional shifts in the angular momentum threshold. The comparison between the two panels demonstrates that the sign of the Symmergent parameter controls the correlation between $r_{isco}$ and $\mathcal{L}_{isco}$. The $\gamma > 0$  branch produces a more spin-separated structure in the ISCO phase space, whereas the $\gamma < 0$  branch leads to a more uniformly increasing angular-momentum profile with radius. Thus, the $(\mathcal{L}_{isco},r_{isco})$ plane provides a useful diagnostic of how symmergent deformation and spin–curvature coupling jointly reorganize the strong field orbital dynamics of spinning particles.

\begin{figure}[h!]
   \centering
   \includegraphics[width=0.475\textwidth]{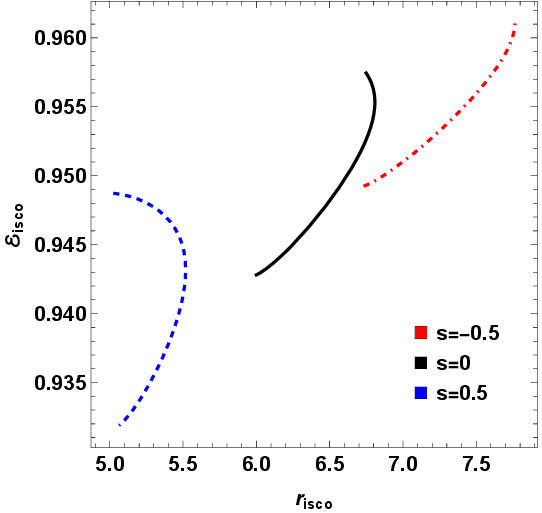}    \includegraphics[width=0.475\textwidth]{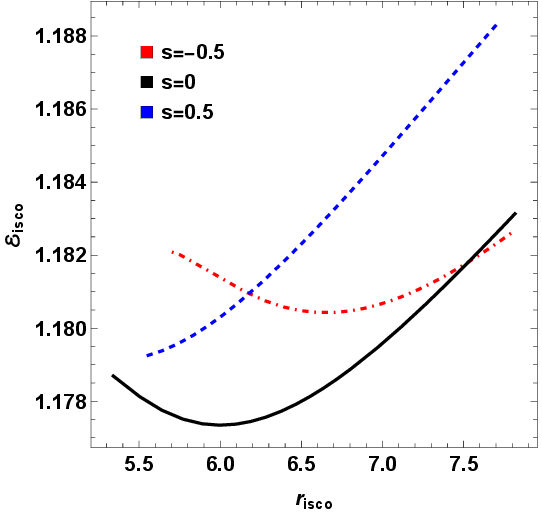}
    \caption{The variation of the energy at the ISCO of spinning particles with respect to $r_{\rm ISCO}$ for negative values of $n_B - n_F < 0$ ($\gamma >0$) (left panel) and positive values of $n_B - n_F > 0$ ($\gamma < 0$) (right panel).}
    \label{Energy-ISCO-spin}
\end{figure}

Fig. \ref{Energy-ISCO-spin} shows the conserved specific energy of spinning particles at the ISCO as a function of the ISCO radius. For the $\gamma>0$ branch, the ISCO energy remains below unity over the displayed range. This indicates that the stable circular orbits of spinning particles remain in a bound energy regime. The spin dependence is clearly resolved: the anti-aligned spin case, $s=-0.5$, gives the largest energy values, the spinless case lies in between, and the aligned spin case, gives the lowest energies. Thus, spin–curvature coupling shifts not only the location of the ISCO but also the energetic threshold required to sustain stability. The positive branch exhibits a qualitatively different behavior. For $\gamma<0$, the ISCO energy lies above unity. This shows that the positive branch drives the stable configurations into a higher energy regime compared with $\gamma>0$. The energy–radius relation also becomes more spin-sensitive: the $s=0$ and $s=-0.5$ curves display shallow minima, while the $s=0.5$ curve increases approximately monotonically over the shown interval. This behavior reflects the stronger influence of the oscillatory symmergent branch on the spinning particle dynamics. The comparison of the two panels demonstrates that the sign of the Bose–Fermi imbalance strongly affects the energetic character of ISCO configurations. While the $\gamma>0$ branch preserves a Schwarzschild-like bound-energy structure, the $\gamma<0$ branch significantly raises the ISCO energy above unity. Therefore, the relation between $\mathcal{E}_{isco}$ and $r_{isco}$ provides a useful diagnostic of the branch dependent strongfield dynamics of spinning particles in symmergent gravity.

\begin{figure}[htb!]
   \centering
    \includegraphics[width=0.47\textwidth]{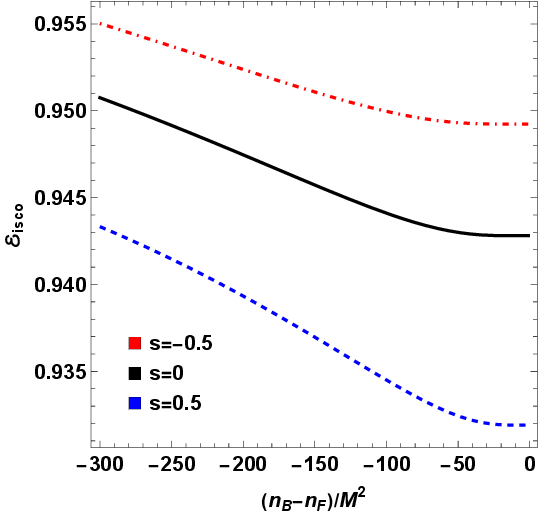}
   \includegraphics[width=0.48\textwidth]{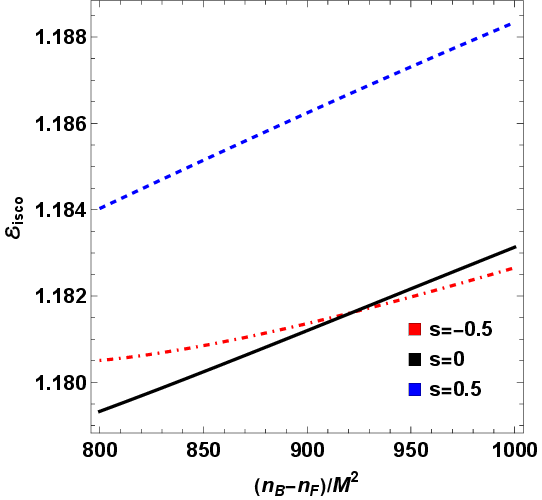}
    \caption{The variation of the conserved energy of spinning particles at the ISCO for negative values of $(n_B - n_F)/M^2$ ($\gamma >0$) (left panel) and positive values of $(n_B - n_F)/M^2$ ($\gamma < 0$) (right panel).}
    \label{Energy-ISCO-spin-nM}
\end{figure}

Fig. \ref{Energy-ISCO-spin-nM} shows the conserved specific energy of spinning particles at the ISCO as a function of the symmergent parameter $(n_B - n_F)/M^2$. In the $\gamma>0$ branch, $\mathcal{E}_{isco}$ decreases as $(n_B - n_F)/M^2$ approaches zero from below. The energy remains below unity throughout the displayed interval, indicating that the ISCO configurations of spinning particles stay in a bound-energy regime. The spin dependence is clearly ordered: the anti-aligned spin case, $s=-0.5$, gives the largest ISCO energy, the spinless case lies in between, and the aligned spin case, $s=0.5$, gives the smallest energy. This shows that spin–curvature coupling systematically shifts the energetic threshold of stable circular motion in the negative branch. The positive branch displays the opposite trend. For $\gamma<0$, $\mathcal{E}_{isco}$ increases with increasing positive $(n_B - n_F)/M^2$, and all curves lie above unity. This indicates that the positive branch drives the spinning-particle ISCO configurations into a higher-energy regime compared with the $\gamma>0$ branch. The spin ordering is also reversed: the $s=0.5$ curve gives the largest energy, whereas the $s=-0.5$ curve gives the lowest values. Hence, the effect of spin orientation on the ISCO energy depends strongly on the sign of the symmergent parameter $(n_B - n_F)/M^2$. The comparison of the two panels shows that the two symmergent branches have qualitatively different energetic behavior. In the negative branch, the deformation preserves a Schwarzschild-like bound-energy structure, while in the positive branch the ISCO energy is shifted above unity and grows with the symmergent parameter. Therefore, the dependence of $\mathcal{E}_{isco}$ on $(n_B - n_F)/M^2$ provides a sensitive diagnostic of branch dependent spinning particle dynamics in Symmergent gravity.

\begin{figure}[h!]
   \centering
 \includegraphics[width=0.48\textwidth]{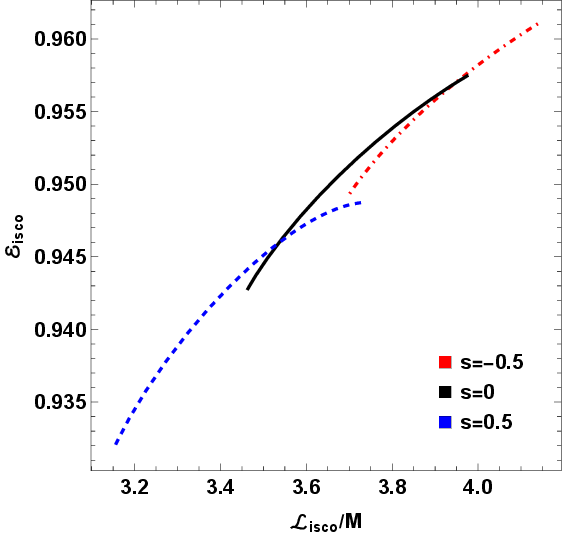}
\includegraphics[width=0.48\textwidth]{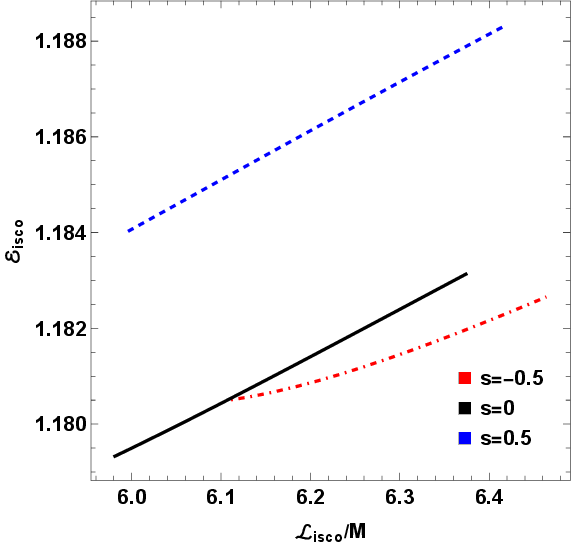}
    \caption{The relation between the specific energy $\mathcal E_{\rm isco}$ and the angular momentum $\mathcal L_{\rm isco}$ at the ISCO for spinning particles for negative values of $n_B - n_F < 0$ ($\gamma >0$) (left panel) and positive values of $n_B - n_F > 0$ ($\gamma < 0$) (right panel).}
    \label{E-L-ISCO-spin}
\end{figure}

Fig. \ref{E-L-ISCO-spin} displays the relation between the conserved specific energy $\mathcal{E}_{isco}$ and the angular momentum $\mathcal{L}_{isco}$ at the ISCO for spinning particles. This parametrically combines the separate behaviors of the ISCO energy and angular momentum into a single diagnostic of the strong-field orbital structure. In the negative branch, $(n_B - n_F) < 0$ ($\gamma >0$), all spin configurations show a clear positive correlation between $\mathcal{E}_{isco}$ and $\mathcal{L}_{isco}$: larger angular momentum at the ISCO is associated with larger conserved energy. The spin dependence shifts the curves systematically. The $s=0.5$ configuration occupies the lower $\mathcal{E}_{isco}$ - lower $\mathcal{L}_{isco}$ region, while the $s=-0.5$ case is displaced toward higher values of both quantities, with the spinless case lying in between. This indicates that, in the $\gamma >0$ branch, spin–curvature coupling jointly reshapes the energy and angular-momentum thresholds of marginally stable circular motion. Since all curves remain in the regime $\mathcal{E}_{isco}<1$, the ISCO configurations in this branch preserve a bound orbit character. In the positive branch, $(n_B - n_F) > 0$ ($\gamma < 0$), the same positive correlation persists, but the entire relation is shifted to higher energy and angular momentum scales. In particular, the curves lie in the regime $\mathcal{E}_{isco}>1$, showing that the positive branch drives the ISCO into a more energetic orbital sector. The $s=0.5$ curve reaches the highest energy values, while the other spin configurations lie below it. Thus, in the $\gamma < 0$ branch, spin not only shifts the ISCO parameters but also enhances the energetic cost of maintaining stable circular orbits. Overall, Fig. \ref{E-L-ISCO-spin} shows that the ISCO energy and angular momentum of spinning particles are not independent quantities, but correlated observables jointly controlled by the symmergent branch structure and the spin orientation. The ($\mathcal{L}_{isco}, \mathcal{E}_{isco}$) plane therefore provides a compact diagnostic of how spin–curvature coupling and symmergent deformation reorganize the strong-field orbital dynamics.

\section{Observational frequency shifts from the asymptotically flat symmergent black-hole}
\label{sec4}

In addition to the motion and collision of massive probes, the asymptotically-flat symmergent black hole can be tested through the observational frequency shift of photons emitted by particles moving on circular geodesics around the black hole using the method in \cite{Martinez-Valera:2023guj,Ovgun:2026max}. This observable is particularly useful because it links directly measurable redshift/blueshift data to the underlying metric functions. In this section, we adapt the general relativistic formalism for observational redshift in static, spherically symmetric spacetimes to the asymptotically-flat symmergent black hole geometry.

\subsection{General formalism for circular emitters}

Let us consider the general static, spherically symmetric line element
\begin{equation}
ds^{2}=-A(r)\,dt^{2}+\frac{dr^{2}}{B(r)}+C(r)\left(d\theta^{2}+\sin^{2}\theta\,d\phi^{2}\right),
\label{metric_redshift_sym}
\end{equation}
and restrict the motion to the equatorial plane $\theta=\pi/2$. For a massive particle on a timelike geodesic, the conserved specific energy and specific angular momentum are
\begin{equation}
\mathcal{E}=A(r)\,\dot t,
\qquad
\mathcal{L}=C(r)\,\dot\phi,
\label{EL_red}
\end{equation}
while the radial equation reads
\begin{equation}
\dot r^{2}=B(r)\left[\frac{\mathcal{E}^{2}}{A(r)}-\left(1+\frac{\mathcal{L}^{2}}{C(r)}\right)\right].
\label{radial_red}
\end{equation}
For circular geodesics, one imposes
\begin{equation}
\dot r=0,
\qquad
\frac{d}{dr}\dot r^{2}=0,
\end{equation}
which yields
\begin{equation}
\mathcal{E}^{2}=\frac{A(r)^{2}C'(r)}{A(r)C'(r)-C(r)A'(r)},
\label{E2_circ_red}
\end{equation}
\begin{equation}
\mathcal{L}^{2}=\frac{C(r)^{2}A'(r)}{A(r)C'(r)-C(r)A'(r)}.
\label{L2_circ_red}
\end{equation}
Hence the nonvanishing components of the emitter four-velocity are
\begin{equation}
U^{t}_{e}=\sqrt{\frac{C'(r_e)}{A(r_e)C'(r_e)-C(r_e)A'(r_e)}},
\label{Ut_general_red}
\end{equation}
\begin{equation}
U^{\phi}_{e}=\sqrt{\frac{A'(r_e)}{A(r_e)C'(r_e)-C(r_e)A'(r_e)}}.
\label{Uphi_general_red}
\end{equation}

Now let $k^\mu$ be the photon wave-vector. The conserved photon energy and angular momentum are
\begin{equation}
\mathcal{E}_{\gamma}=A(r)\,k^{t},
\qquad
\mathcal{L}_{\gamma}=C(r)\,k^{\phi}.
\label{photon_consts_red}
\end{equation}
At the diametrically opposite points where the radial component of the photon vanishes, $k^{r}=0$, the null condition implies that the photon impact parameter
\begin{equation}
b_{\gamma}\equiv \frac{\mathcal{L}_{\gamma}}{\mathcal{E}_{\gamma}}
\end{equation}
takes the values
\begin{equation}
b_{\gamma\pm}=\pm \sqrt{\frac{C(r)}{A(r)}}.
\label{impact_red}
\end{equation}

The photon frequency measured by an observer with four-velocity $U^\mu$ is
\begin{equation}
\omega=-k_{\mu}U^{\mu}.
\end{equation}
For a static observer at infinity, one has $U^\mu_{d}=(1,0,0,0)$ because asymptotic flatness implies $A(r)\to 1$ as $r\to\infty$. Therefore, the total observational frequency shift is
\begin{equation}
1+z_{\pm}=\frac{\omega_{e}}{\omega_{d}}
=
U^{t}_{e}\mp b_{\gamma\pm}U^{\phi}_{e},
\label{z_def_general}
\end{equation}
which can be written explicitly as
\begin{equation}
1+z_{\pm}
=
\frac{\sqrt{C'(r_e)}\pm \sqrt{\dfrac{C(r_e)A'(r_e)}{A(r_e)}}}
{\sqrt{A(r_e)C'(r_e)-C(r_e)A'(r_e)}}.
\label{z_general_metric}
\end{equation}
Here $z_{+}$ and $z_{-}$ denote the two frequency-shift branches associated with photons emitted from the receding and approaching sides of the circular source, respectively. The approaching-side branch represents an actual blueshift only where $z_{-}<0$; it may remain positive in part of the parameter space, in which case the approaching photon is still redshifted. The sign of $z_{-}$ must therefore be retained rather than plotting only $|z_{-}|$. It is also useful to separate the total shift into a gravitational and a kinematic part,
\begin{equation}
z_{\pm}=z_{g}+z_{{\rm kin}\,\pm},
\end{equation}
with
\begin{equation}
1+z_{g}=\frac{\sqrt{C'(r_e)}}{\sqrt{A(r_e)C'(r_e)-C(r_e)A'(r_e)}},
\label{zg_general_metric}
\end{equation}
and
\begin{equation}
z_{{\rm kin}\,\pm}
=
\pm
\frac{\sqrt{\dfrac{C(r_e)A'(r_e)}{A(r_e)}}}
{\sqrt{A(r_e)C'(r_e)-C(r_e)A'(r_e)}}.
\label{zkin_general_metric}
\end{equation}

Although the notation $z_g$ is retained for convenience, the factor $1+z_g=U_e^t$ is not a purely gravitational contribution. It also contains the transverse kinematic time dilation of the circularly moving emitter. Consequently, a negative value of $z_g$ should not be described as an unambiguous purely gravitational blueshift.

A useful model-independent identity follows immediately from Eq.~(\ref{z_general_metric}):
\begin{equation}
(1+z_{+})(1+z_{-})=\frac{1}{A(r_e)}.
\label{product_general_red}
\end{equation}
Thus the product of the observable redshift and blueshift directly reconstructs the lapse function at the emitter radius.

\subsection{Application to the symmergent black-hole exterior}

For the perturbative symmergent exterior, we set $\varphi(r)=\epsilon f(r)$ and write
\begin{align}
ds^2
=
e^{-\epsilon f(r)}
\left(
-\Psi(r)dt^2
+
\frac{dr^2}{\Psi(r)}
+
r^2d\Omega^2
\right)
+
\mathcal O(\epsilon^2),
\qquad
\Psi(r)=1-\frac{2M}{r}.
\label{metric_sym_red}
\end{align}
Thus,
\begin{align}
A(r)&=e^{-\epsilon f(r)}\Psi(r)+\mathcal O(\epsilon^2),
\nonumber\\
B(r)&=e^{\epsilon f(r)}\Psi(r)+\mathcal O(\epsilon^2),
\nonumber\\
C(r)&=e^{-\epsilon f(r)}r^2+\mathcal O(\epsilon^2).
\label{ABC_sym_red}
\end{align}
The function $f(r)$ satisfies
\begin{align}
\left(r^2\Psi(r)f'(r)\right)'
=
\gamma r^2f(r),
\qquad
\gamma=-\frac{64\pi}{3(n_B-n_F)}.
\label{phi_eq_red}
\end{align}
All formulas in this subsection are meaningful only up to $\mathcal O(\epsilon)$ and only where $|\epsilon f(r)|\leq\eta$.

Using Eq.~\eqref{ABC_sym_red}, one obtains
\begin{align}
A'(r)
&=
e^{-\epsilon f(r)}\left[\Psi'(r)-\epsilon\Psi(r)f'(r)\right]
+
\mathcal O(\epsilon^2),
\nonumber\\
C'(r)
&=
e^{-\epsilon f(r)}\left[2r-\epsilon r^2f'(r)\right]
+
\mathcal O(\epsilon^2).
\label{AprimeCprime_sym_red}
\end{align}
The common circular-orbit denominator simplifies to
\begin{align}
A(r)C'(r)-C(r)A'(r)
=
2e^{-2\epsilon f(r)}(r-3M)
+
\mathcal O(\epsilon^2).
\label{ACmCAprime_sym}
\end{align}
Accordingly,
\begin{align}
U^t_e
&=
e^{\epsilon f(r_e)/2}
\sqrt{
\frac{2r_e-\epsilon r_e^2f'(r_e)}
{2(r_e-3M)}
}
+
\mathcal O(\epsilon^2),
\label{Ut_sym_red}
\\
U^\phi_e
&=
e^{\epsilon f(r_e)/2}
\sqrt{
\frac{\dfrac{2M}{r_e^2}-\epsilon\Psi(r_e)f'(r_e)}
{2(r_e-3M)}
}
+
\mathcal O(\epsilon^2).
\label{Uphi_sym_red}
\end{align}
The photon impact parameter remains conformally invariant to the working order,
\begin{align}
b_{\gamma\pm}
=
\pm\sqrt{\frac{C(r_e)}{A(r_e)}}
=
\pm\frac{r_e}{\sqrt{\Psi(r_e)}}
+
\mathcal O(\epsilon^2).
\label{impact_sym_red}
\end{align}
The two observational frequency-shift branches are therefore
\begin{align}
1+z_\pm
=
\frac{e^{\epsilon f(r_e)/2}}
{\sqrt{2(r_e-3M)}}
\left[
\sqrt{2r_e-\epsilon r_e^2f'(r_e)}
\pm
\sqrt{
\frac{2Mr_e}{r_e-2M}
-
\epsilon r_e^2f'(r_e)
}
\right]
+
\mathcal O(\epsilon^2).
\label{z_sym_final}
\end{align}
The time-dilation and signed kinematic pieces are
\begin{align}
1+z_g
&=
\frac{e^{\epsilon f(r_e)/2}}
{\sqrt{2(r_e-3M)}}
\sqrt{2r_e-\epsilon r_e^2f'(r_e)}
+
\mathcal O(\epsilon^2),
\label{zg_sym}
\\
z_{{\rm kin}\,\pm}
&=
\pm
\frac{e^{\epsilon f(r_e)/2}}
{\sqrt{2(r_e-3M)}}
\sqrt{
\frac{2Mr_e}{r_e-2M}
-
\epsilon r_e^2f'(r_e)
}
+
\mathcal O(\epsilon^2).
\label{zkin_sym}
\end{align}
Here $z_g$ includes both gravitational and transverse kinematic time dilation and should not be interpreted as a purely gravitational shift.

The product identity becomes
\begin{align}
(1+z_+)(1+z_-)
=
\frac{1}{A(r_e)}
=
\frac{e^{\epsilon f(r_e)}}{1-2M/r_e}
+
\mathcal O(\epsilon^2).
\label{product_sym}
\end{align}
Under the assumptions that $M$ and $r_e$ are independently known and that both shifts arise from the same circular ring, the conformal deformation may be reconstructed as
\begin{align}
\epsilon f(r_e)
=
\ln\!\left[
(1+z_+)(1+z_-)
\left(1-\frac{2M}{r_e}\right)
\right]
+
\mathcal O(\epsilon^2).
\label{phi-reconstruction-redshift}
\end{align}
Equivalently, if the conformal deformation is supplied independently, the mass relation is
\begin{align}
M
=
\frac{r_e}{2}
\left[
1-
\frac{e^{\epsilon f(r_e)}}{(1+z_+)(1+z_-)}
\right]
+
\mathcal O(\epsilon^2).
\label{mass_from_redshift_sym}
\end{align}
The reconstruction does not by itself uniquely determine $n_B-n_F$, because it is degenerate with the independent amplitude $\epsilon$ and, for $\gamma<0$, with the phase $\delta$.

At large radius,
\begin{align}
f(r)
&\simeq
\frac{e^{-\sqrt{\gamma}\,r}}{\sqrt{\gamma}\,r},
&&\gamma>0,
\label{phi_asym_pos_red}
\\
f(r)
&\simeq
\frac{\cos\!\left(\sqrt{|\gamma|}\,r+\delta\right)}
{\sqrt{|\gamma|}\,r},
&&\gamma<0.
\label{phi_asym_neg_red}
\end{align}
Thus the $\gamma>0$ correction is short-ranged, whereas the $\gamma<0$ correction has an oscillatory inverse-radius envelope. These asymptotic patterns can help distinguish the two branches, but a quantitative inference of the microscopic particle content requires independent control of $\epsilon$, $\delta$, $M$, and $r_e$.

In the Schwarzschild limit $\epsilon f\rightarrow0$, one recovers
\begin{align}
1+z_\pm^{\rm Schw}
=
\frac{1}{\sqrt{2(r_e-3M)}}
\left[
\sqrt{2r_e}
\pm
\sqrt{\frac{2Mr_e}{r_e-2M}}
\right],
\label{z_schw_limit}
\end{align}
and
\begin{align}
(1+z_+^{\rm Schw})(1+z_-^{\rm Schw})
=
\frac{1}{1-2M/r_e}.
\label{product_schw_limit}
\end{align}

\subsection{Numerical results and discussion}
\label{sec:redshift-numerical}

To evaluate the frequency-shift observables consistently, we solve Eq.~\eqref{phi_eq_red} for the normalized function $f(r)$ rather than absorbing its amplitude into the solution. Introducing $v(r)=f'(r)$ gives
\begin{align}
f'(r)&=v(r),
\nonumber\\
v'(r)
&=
\frac{\gamma f(r)}{\Psi(r)}
-
\frac{2v(r)}{r}
-
\frac{2Mv(r)}{r^2\Psi(r)}.
\label{phi-system}
\end{align}
The system is integrated inward from
\begin{align}
r_{\rm far}
=
\max\!\left(100M,\frac{20}{\sqrt{|\gamma|}}\right),
\end{align}
using Eqs.~\eqref{phi_asym_pos_red} and \eqref{phi_asym_neg_red} and their derivatives as Cauchy data. For the illustrative comparison we set $\delta=0$ and use a common amplitude $\epsilon=0.05$. The numerical integration should use a relative and absolute tolerance of at least $10^{-10}$ and a maximum radial step small compared with the oscillation length $2\pi/\sqrt{|\gamma|}$. Every plotted point is filtered by
\begin{align}
|\epsilon f(r)|&\leq0.1,
\nonumber\\
\mathcal E_c^2&>0,
\qquad
\mathcal L_c^2\geq0,
\qquad
\dot t>0.
\label{redshift-numerical-filter}
\end{align}
When a curve is described as stable, we additionally require $\partial_r^2V_{\rm eff}>0$. The algebraic product identity in Eq.~\eqref{product_sym} is used as an internal implementation check, but agreement with that identity is not by itself an independent validation of the ODE solution.

The parameter values $|n_B-n_F|\in\{500,1000,2000,5000\}$ correspond to
$|\gamma|\in[1.34\times10^{-2},1.34\times10^{-1}]$. All curves shown below
are generated with the explicit common amplitude $\epsilon=0.05$ and phase
$\delta=0$, integrating Eq.~\eqref{phi-system} inward from $r_{\rm far}$
with relative tolerance $10^{-11}$ and a maximum step small compared with
$2\pi/\sqrt{|\gamma|}$. For every curve we monitor
$\eta_{\rm max}=\max_{r_e\in[6M,30M]}|\epsilon f(r_e)|$ and find
$\eta_{\rm max}\leq0.078$, attained by the $|n_B-n_F|=5000$ oscillatory
configuration; the perturbative bound $|\epsilon f|\leq0.1$ is therefore
respected on the complete displayed interval and no curve is removed. Two
implementation checks are used throughout: the product identity
Eq.~\eqref{product_sym} is satisfied to machine precision
($\sim10^{-15}$), and the $\epsilon\to0$ limit of the numerical pipeline
reproduces the Schwarzschild expressions Eq.~\eqref{z_schw_limit} exactly.
We also note that for the conformal metric functions
Eq.~\eqref{ABC_sym_red} the circular-orbit denominator obeys
$A(r)C'(r)-C(r)A'(r)=2e^{-2\epsilon f(r)}(r-3M)$ \emph{exactly}, since the
$f'$ contributions cancel; the positivity condition in
Eq.~\eqref{neutral-circular-conditions} therefore reduces precisely to
$r_e>3M$, and the nontrivial content of the filter
Eq.~\eqref{redshift-numerical-filter} resides in the conditions
$\mathcal L_c^2\geq0$ and $|\epsilon f|\leq\eta$.

\begin{figure}[htb!]
   \centering
    \includegraphics[width=1\textwidth]{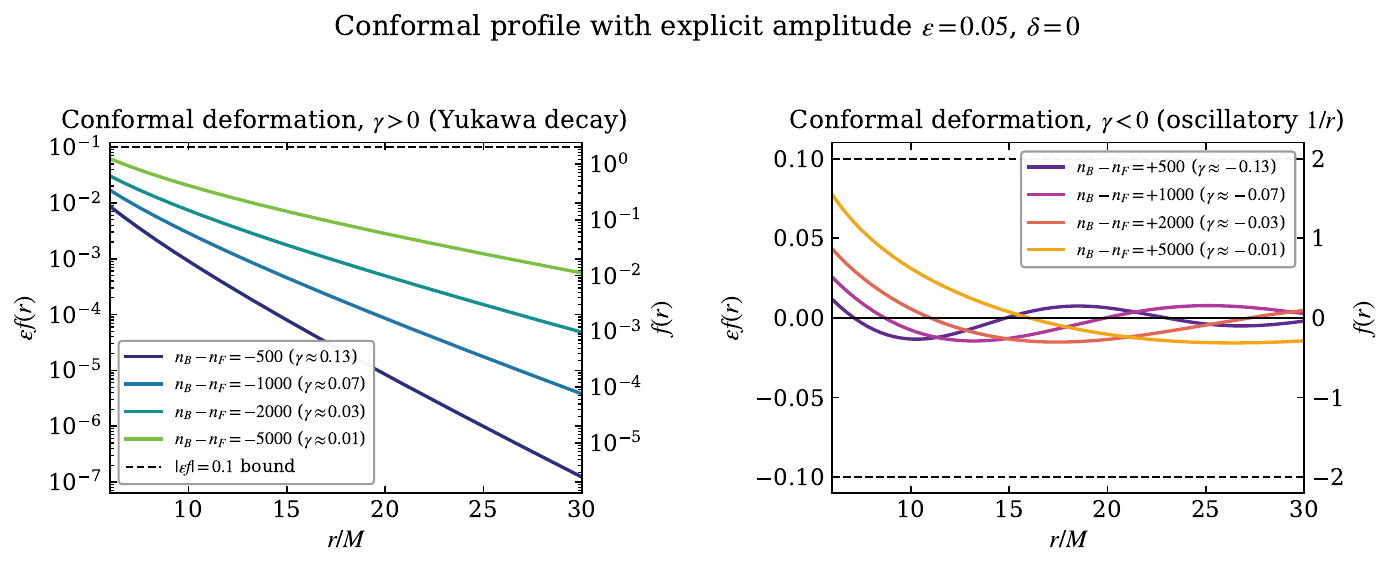}
    \caption{Physical conformal deformation
    $\epsilon f(r)$ obtained by integrating Eq.~\eqref{phi-system} inward
    from $r_{\rm far}$ with the explicit common amplitude $\epsilon=0.05$
    and phase $\delta=0$; the normalized profile $f(r)$ is readable from
    the secondary axis. Left: Yukawa branch $\gamma>0$ ($n_B<n_F$),
    logarithmic scale. Right: oscillatory branch $\gamma<0$ ($n_B>n_F$).
    Dashed lines mark the perturbative bound $|\epsilon f|=0.1$; all
    curves satisfy it on the complete interval, with
    $\eta_{\rm max}\leq0.078$, so no curve is truncated or removed.}
    \label{fig:phi-num}
\end{figure}

Figure~\ref{fig:phi-num} displays the physical deformation
$\epsilon f(r)$ obtained from the inward integration, together with the
normalized profile $f(r)=\epsilon f(r)/\epsilon$ readable from the
secondary axis. The $\gamma>0$ branch decays monotonically on the Yukawa
scale $1/\sqrt{\gamma}$, so that larger $|\gamma|$ (smaller $|n_B-n_F|$)
produces a faster falloff and a smaller residual deformation at fixed
radius. The $\gamma<0$ branch exhibits the oscillatory $1/r$ envelope,
whose wavelength $2\pi/\sqrt{|\gamma|}$ grows with $|n_B-n_F|$; for
$|n_B-n_F|=5000$ less than one full oscillation fits inside the displayed
window, whereas for $|n_B-n_F|=500$ several sign changes occur. In both
branches the largest deformation is reached at the inner edge $r_e=6M$ and
remains below the bound $|\epsilon f|=0.1$ (dashed lines). The oscillatory
profile also depends on the phase $\delta$, so the locations of its
extrema and zero crossings are not determined by $\gamma$ alone.

\begin{figure}[htb!]
   \centering
    \includegraphics[width=1\textwidth]{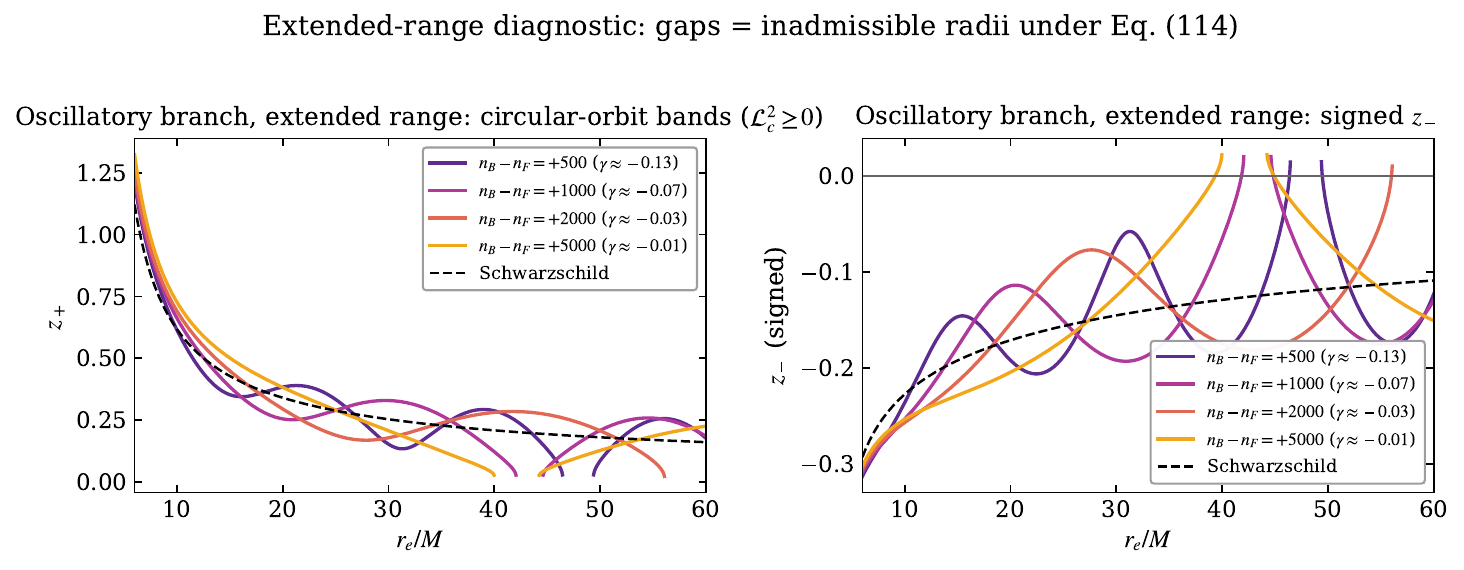}
    \caption{Extended-range diagnostic for the
    oscillatory branch, $r_e/M\in[6,60]$, with $\epsilon=0.05$ and
    $\delta=0$: signed $z_+$ (left) and signed $z_-$ (right). The gaps in
    the curves are radial intervals excluded by the filter
    Eq.~\eqref{redshift-numerical-filter}, where $\mathcal L_c^2<0$ and no
    circular timelike geodesic exists; the first such intervals open at
    $r_e/M\simeq40$--$56$ depending on $|n_B-n_F|$. The admissible
    segments are bands of circular-orbit candidates whose stability must
    be established separately through $\partial_r^2V_{\rm eff}$. The
    number and location of the bands depend on $\epsilon$, $\delta$, and
    $\gamma$ jointly.}
    \label{fig:bands}
\end{figure}

\medskip

\begin{figure}[htb!]
   \centering
    \includegraphics[width=1\textwidth]{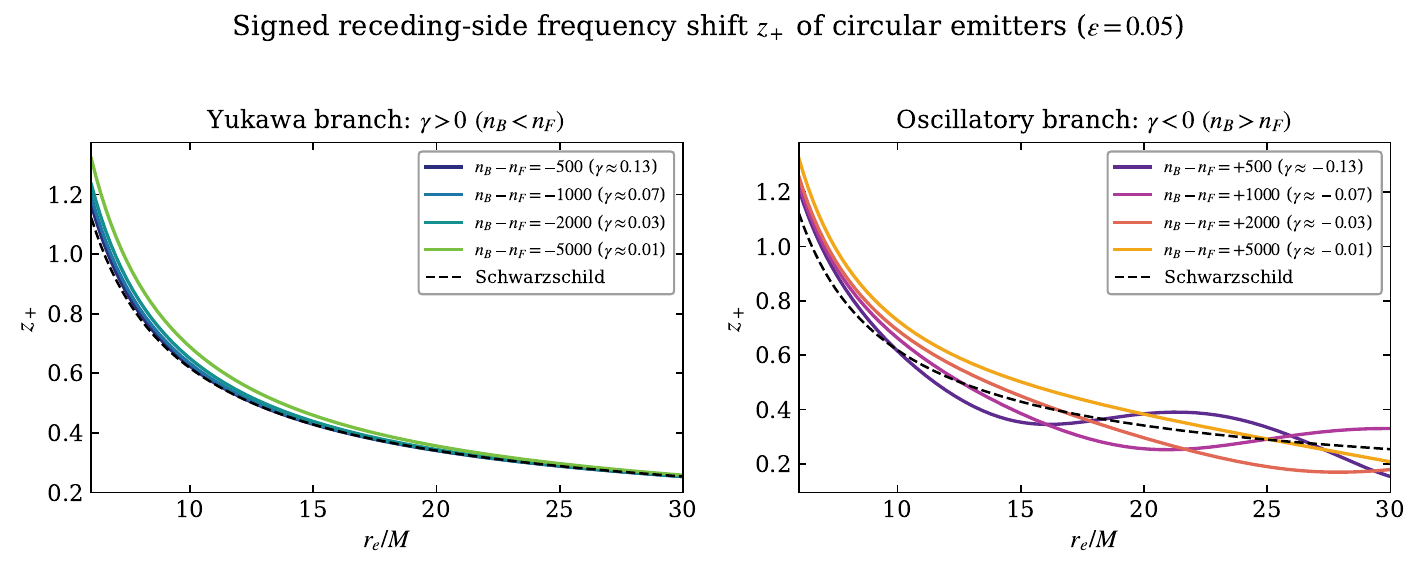}
    \caption{Signed receding-side frequency shift
    $z_+(r_e)$ of circular geodesic emitters for $\epsilon=0.05$,
    $\delta=0$, compared with the Schwarzschild reference (dashed). Left:
    Yukawa branch, showing a smooth, strictly positive, short-ranged
    enhancement that is largest at the inner edge and decays outward.
    Right: oscillatory branch, showing modulations that alternate around
    the Schwarzschild curve with wavelength $2\pi/\sqrt{|\gamma|}$. All
    points satisfy the filter Eq.~\eqref{redshift-numerical-filter}; at
    this amplitude the filter admits the entire interval
    $r_e/M\in[6,30]$, and the disconnected circular-orbit bands of the
    oscillatory branch open only at $r_e/M\gtrsim40$
    (Fig.~\ref{fig:bands}). Plotted radii are circular-orbit candidates
    and are not implied to be stable unless $\partial_r^2V_{\rm eff}>0$ is
    verified separately.}
    \label{fig:zplus}
\end{figure}

\noindent\textit{Receding-side frequency shift.}
Figure~\ref{fig:zplus} displays the signed branch $z_+(r_e)$. In the
Yukawa branch, $f(r)>0$ throughout the window, so all curves lie above the
Schwarzschild reference; the excess is largest at the inner edge and
decays smoothly outward, with the slowest decay for the smallest
$|\gamma|$. In the oscillatory branch, $z_+$ alternates around the
Schwarzschild curve following the sign of the local conformal
deformation, and the crossings track the zeros of $\epsilon f$ and
$f'$. The overall deviations are at the few-percent level, consistent
with $\eta_{\rm max}\leq0.078$. We emphasize that, at the amplitude
$\epsilon=0.05$ adopted here, the filter
Eq.~\eqref{redshift-numerical-filter} admits the \emph{entire} interval
$r_e/M\in[6,30]$: the disconnected circular-orbit bands of the oscillatory
branch open up only at larger radii, $r_e/M\gtrsim40$, where the intervals
with $\mathcal L_c^2<0$ first appear because the Schwarzschild term
$2M/r^2$ in $A'(r)$ falls off faster than the oscillatory contribution
$\epsilon\,\Psi f'\sim\epsilon/r$ (see
Fig.~\ref{fig:bands}). The existence of gaps and their locations therefore
depend jointly on $\epsilon$, $\delta$, and $\gamma$, and are not intrinsic
to the plotted window. Circular-orbit candidates shown here are not
automatically stable: evaluating $\partial_r^2V_{\rm eff}$ separately, we
find that the Yukawa branch is stable at essentially all admissible radii,
whereas the oscillatory branch alternates between stable and unstable
radial intervals, with only roughly one half to two thirds of the
admissible radii supporting stable circular motion.

\medskip

\begin{figure}[htb!]
   \centering
    \includegraphics[width=1\textwidth]{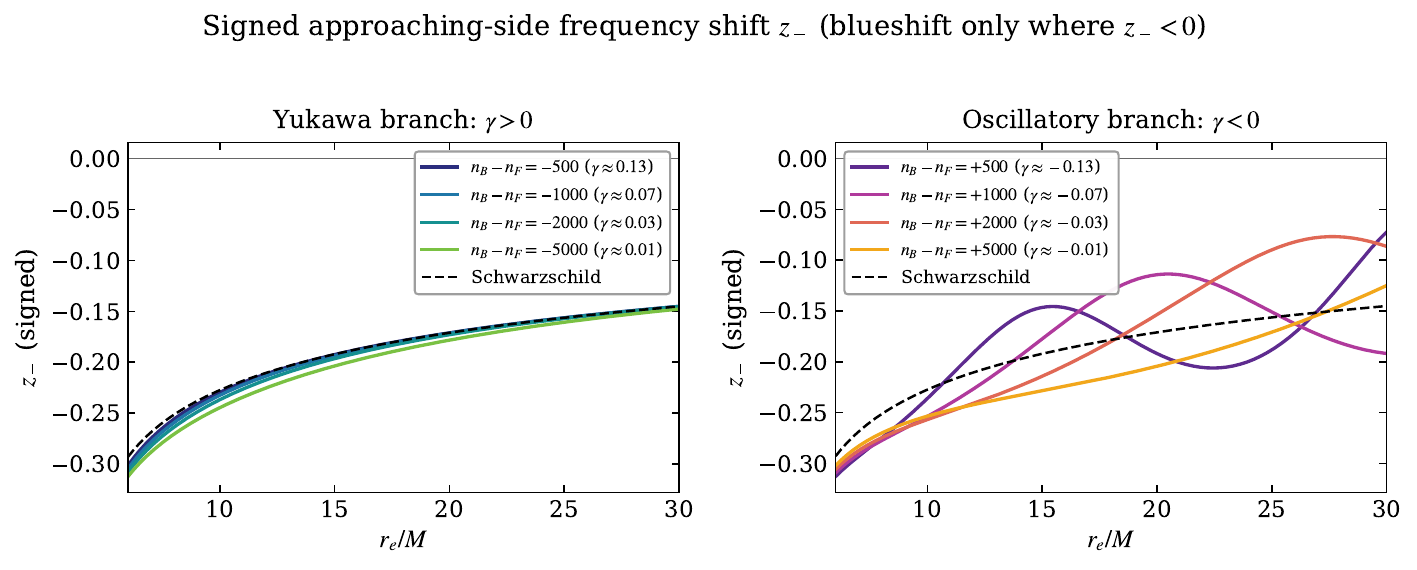}
    \caption{Signed approaching-side frequency shift
    $z_-(r_e)$ for $\epsilon=0.05$, $\delta=0$. A true blueshift
    corresponds to $z_-<0$; for all parameter sets shown, $z_-$ remains
    strictly negative over $r_e/M\in[6,30]$, reaching $z_-\simeq-0.31$ at
    the inner edge. The Yukawa branch (left) smoothly reduces the
    blueshift magnitude at small radii relative to Schwarzschild
    (dashed), whereas the oscillatory branch (right) alternates around the
    Schwarzschild curve. Zeros of the kinematic contribution satisfy
    $A'(r_e)=0$ and should not be identified with marginally stable
    circular orbits; no such zero occurs inside the displayed window at
    this amplitude.}
    \label{fig:zminus}
\end{figure}

\noindent\textit{Approaching-side frequency shift.}
The signed approaching-side branch $z_-(r_e)$ is shown in
Fig.~\ref{fig:zminus}. For all parameter sets considered, $z_-$ remains
strictly negative over the complete interval $r_e/M\in[6,30]$: the
approaching photon is genuinely blueshifted everywhere in the displayed
window, with the strongest blueshift, $z_-\simeq-0.31$, at the inner edge.
In the Yukawa branch the deviation from Schwarzschild is a smooth,
short-ranged reduction of the blueshift magnitude at small radii, whereas
in the oscillatory branch $z_-$ oscillates around the Schwarzschild
reference and the deviation changes sign along the radial direction.

The kinematic term vanishes when
\begin{align}
\frac{2Mr}{r-2M}-\epsilon r^2f'(r)=0.
\label{zero-kinematic-condition}
\end{align}
Using the metric functions, this condition is equivalent to $A'(r)=0$.
Provided the remaining denominators are regular, it implies
\begin{align}
\Omega^2=\frac{A'(r)}{C'(r)}=0,
\qquad
\mathcal L_c^2=0.
\end{align}
It is not, in general, the marginal-stability or ISCO condition. At such a
radius, $z_+=z_-=z_g$, and the total shift does not generally vanish. For
the amplitude $\epsilon=0.05$ used here, no root of
Eq.~\eqref{zero-kinematic-condition} occurs inside $r_e/M\in[6,30]$; the
first such radii coincide with the inner boundaries of the
$\mathcal L_c^2<0$ gaps at $r_e/M\simeq40$--$56$ visible in
Fig.~\ref{fig:bands}.

\medskip

\begin{figure}[htb!]
   \centering
    \includegraphics[width=1\textwidth]{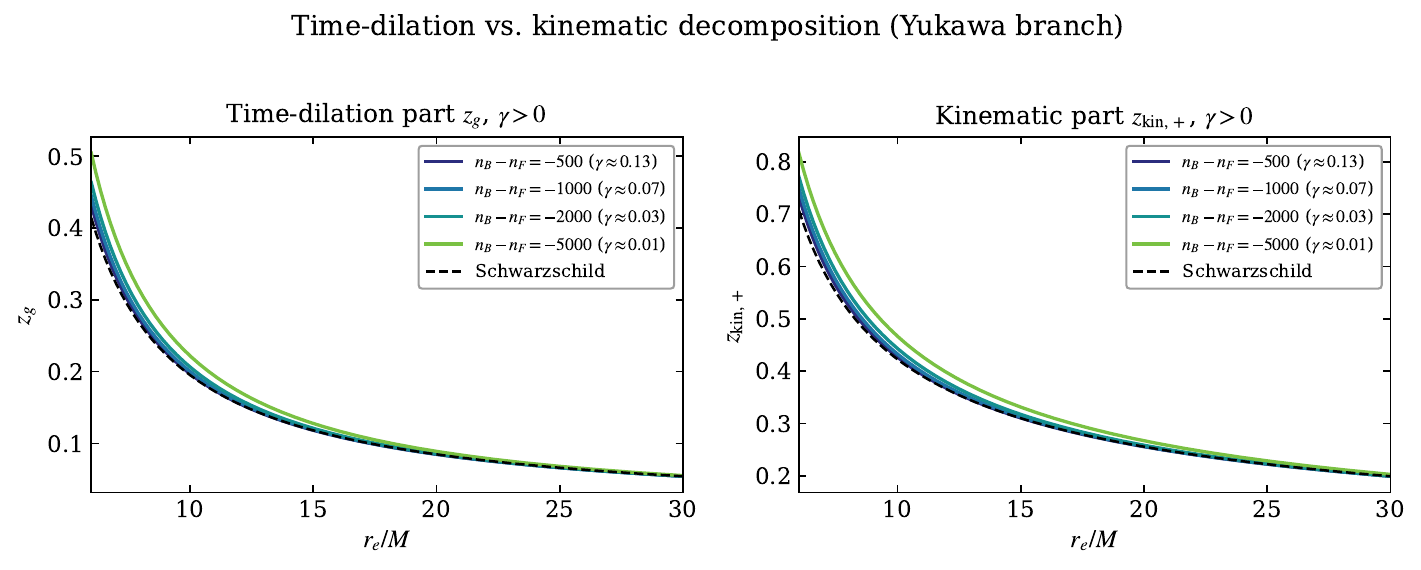}
    \caption{Decomposition of the receding-side shift
    into the time-dilation contribution $z_g$ (left) and the signed
    kinematic contribution $z_{{\rm kin},+}$ (right) for the Yukawa branch
    with $\epsilon=0.05$, compared with Schwarzschild (dashed). Both
    pieces are smooth, positive, and enhanced by the positive conformal
    deformation, most strongly at the inner edge. The quantity $z_g$
    contains both gravitational and transverse kinematic time dilation and
    should not be interpreted as purely gravitational.}
    \label{fig:decomp-pos}
\end{figure}

\begin{figure}[htb!]
   \centering
    \includegraphics[width=1\textwidth]{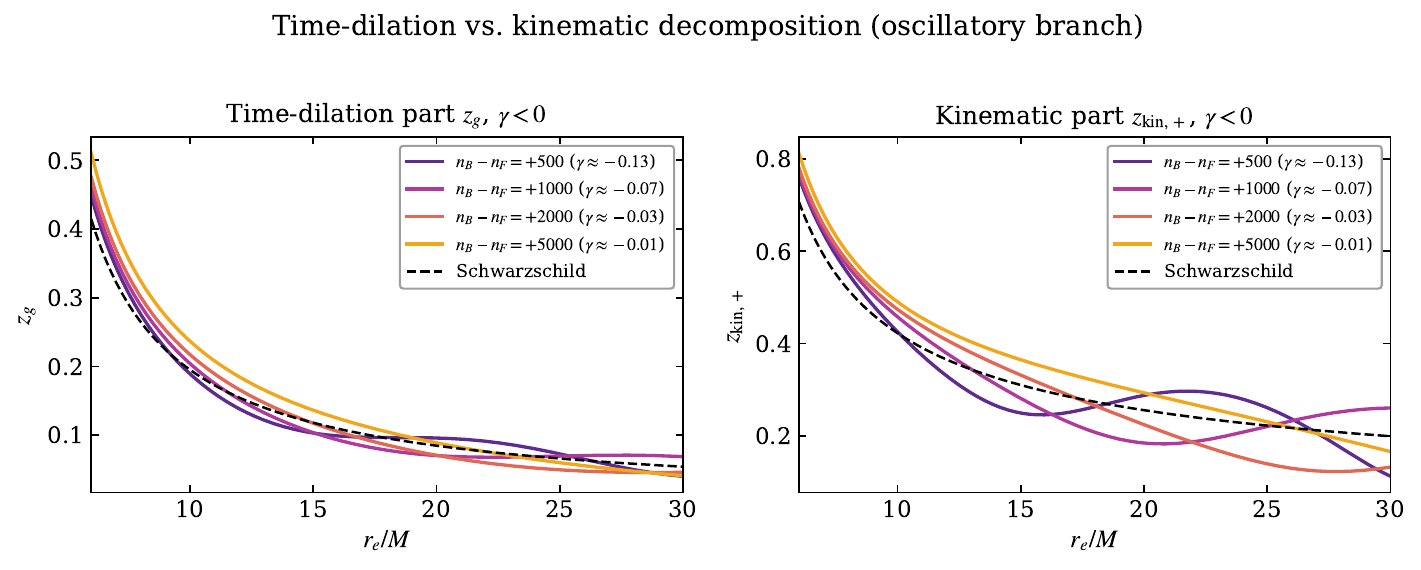}
    \caption{Same decomposition as in
    Fig.~\ref{fig:decomp-pos} for the oscillatory branch. Both pieces
    remain positive over the displayed interval but carry radial
    modulations that alternate around the Schwarzschild reference,
    tracking the local sign of $\epsilon f$ and $\epsilon f'$. The
    time-dilation term $z_g$ contains both gravitational and transverse
    kinematic effects.}
    \label{fig:decomp-neg}
\end{figure}

\noindent\textit{Time-dilation and kinematic decomposition.}
Figures~\ref{fig:decomp-pos} and \ref{fig:decomp-neg} display $z_g$ and
$z_{{\rm kin},+}$ separately. The quantity $z_g$ is a time-dilation
contribution containing both gravitational and transverse kinematic
effects; it is therefore not appropriate to interpret $z_g<0$ as an
unambiguous purely gravitational blueshift. For the present parameter
sets, both $z_g$ and $z_{{\rm kin},+}$ remain positive over the full
window, decreasing from $z_g\simeq0.42$ and
$z_{{\rm kin},+}\simeq0.71$ at $r_e=6M$ toward their asymptotic falloff.
In the Yukawa branch both pieces are smooth, monotonically enhanced
relative to Schwarzschild by the positive conformal deformation. In the
oscillatory branch the two pieces carry radial modulations that are
anticorrelated with the local sign of $\epsilon f$ and $\epsilon f'$;
their relative magnitudes alternate with radius, but no sign change of
either piece occurs inside the displayed interval. All interpretations
remain conditional on the circular-orbit and perturbative filters.

\medskip

\begin{figure}[htb!]
   \centering
    \includegraphics[width=1\textwidth]{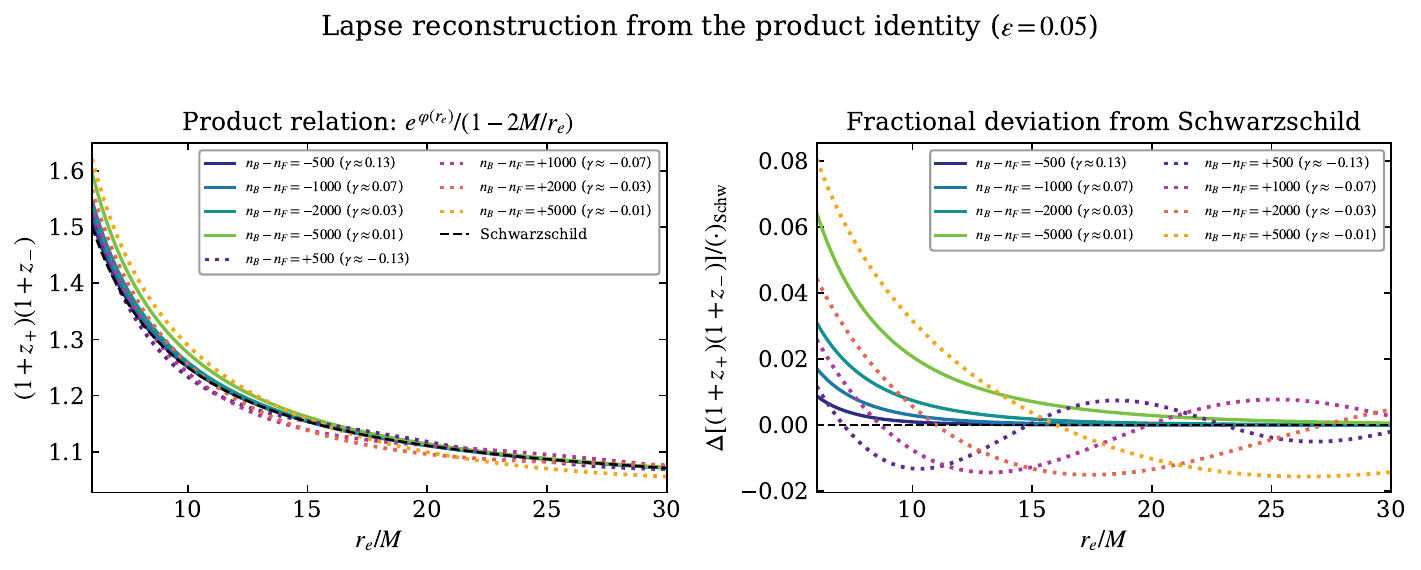}
    \caption{Left: product $(1+z_+)(1+z_-)$, which
    reconstructs $1/A(r_e)=e^{\epsilon f(r_e)}/(1-2M/r_e)$, compared with
    the Schwarzschild result (dashed); solid curves denote the Yukawa
    branch and dotted curves the oscillatory branch. Right: fractional
    deviation of the reconstructed lapse from Schwarzschild, equal to
    $e^{\epsilon f(r_e)}-1$, reaching about $8\%$ at the inner edge for
    $|n_B-n_F|=5000$. The Yukawa branch produces a strictly positive,
    monotonically decaying deviation, whereas the oscillatory branch
    alternates in sign around zero; the spatial pattern, rather than the
    magnitude, distinguishes the sign of $n_B-n_F$. Quantitative inference
    of $|n_B-n_F|$ requires independent knowledge of $\epsilon$, $\delta$,
    $M$, and $r_e$.}
    \label{fig:product}
\end{figure}

\noindent\textit{Product relation and lapse reconstruction.}
The product in Eq.~\eqref{product_sym} reconstructs the lapse at the
emitter radius,
\begin{align}
A(r_e)=\frac{1}{(1+z_+)(1+z_-)}.
\end{align}
The right panel of Fig.~\ref{fig:product} shows the fractional deviation
of the reconstructed lapse from Schwarzschild, which for the conformal
geometry equals $e^{\epsilon f(r_e)}-1$ identically; we verify this
relation to machine precision as an internal consistency check. The
deviation reaches about $8\%$ at the inner edge for $|n_B-n_F|=5000$ and
decreases with $|n_B-n_F|$. The two branches are distinguished by the
\emph{spatial pattern} of the deviation rather than by its magnitude: the
Yukawa branch produces a strictly positive, monotonically decaying
deviation, whereas the oscillatory branch alternates in sign around the
Schwarzschild value with the wavelength $2\pi/\sqrt{|\gamma|}$. A
reconstructed lapse of the former type would indicate $n_B<n_F$, and of
the latter type $n_B>n_F$. Assuming that $M$ and $r_e$ are independently
known, Eq.~\eqref{phi-reconstruction-redshift} then reconstructs the local
conformal deformation; this does not uniquely determine $n_B-n_F$, because
the result is degenerate with the amplitude $\epsilon$, the oscillatory
phase $\delta$, and observational uncertainties in the source radius and
mass. The method also assumes that $z_+$ and $z_-$ arise from the same
circular ring and neglects inclination, light bending, and extended-source
effects.

\medskip

For every numerical curve, we evaluate
\begin{align}
\eta_{\rm max}
=
\max_{r\in[r_{\min},r_{\max}]}
|\epsilon f(r)|.
\end{align}
A curve is retained only if $\eta_{\rm max}\leq0.1$ over its complete displayed interval. Curves violating this condition are removed rather than partially interpreted. The smallness of $|\gamma|$ does not guarantee the validity of the perturbative solution; the relevant expansion parameter is $|\epsilon f(r)|$.

\noindent\textit{Summary of numerical findings.}
The two signs of $\gamma$ lead to qualitatively different spatial
patterns: a smooth, strictly positive, short-ranged deformation of all
frequency-shift observables for $\gamma>0$, and an oscillatory deformation
alternating around the Schwarzschild reference for $\gamma<0$. The sign of
$n_B-n_F$ may therefore be constrained through the spatial form of a
reconstructed lapse, but the magnitude of the microscopic imbalance cannot
be extracted from the frequency shifts alone without controlling the
deformation amplitude, phase, mass, emitter radius, and propagation
effects. At the perturbatively consistent amplitude $\epsilon=0.05$, all
radii in $r_e/M\in[6,30]$ support circular-orbit candidates; disconnected
bands appear only farther out, at $r_e/M\gtrsim40$, and their stability
must in any case be checked separately through
$\partial_r^2V_{\rm eff}$, which reveals alternating stable and unstable
intervals in the oscillatory branch.

\section{Conclusion}
\label{sec5}

We have investigated the dynamics of neutral, electrically charged, and spinning test particles, together with collision energetics and photon frequency shifts, in the perturbative variable-curvature exterior of an asymptotically flat symmergent black hole. At linear order, the geometry is conformally related to the Schwarzschild spacetime through a radial deformation $\varphi(r)$ satisfying a linear field equation. The two signs of the symmergent parameter $\gamma$, determined by the boson--fermion imbalance of the underlying particle spectrum, lead to qualitatively distinct radial profiles. The $\gamma>0$ branch is characterized by a Yukawa-suppressed deformation, whereas the $\gamma<0$ branch exhibits an oscillatory inverse-radius behavior. The overall amplitude of the deformation, and the phase in the oscillatory branch, are integration data fixed by boundary conditions and cannot be inferred from $\gamma$ alone.

For neutral massive particles, we derived the radial equation, effective potential, circular-orbit energy and angular momentum, marginal-stability condition, and center-of-mass energy of particle collisions. The Yukawa branch produces smooth and short-ranged departures from Schwarzschild dynamics. By contrast, the oscillatory branch can generate several disconnected radial intervals in which real circular-orbit solutions exist. These intervals should not automatically be identified with stable orbital regions: stability must be determined independently from the sign of the second radial derivative of the effective potential. When several marginal-stability radii occur, the physical ISCO is identified as the innermost boundary of the stable circular-orbit branch continuously connected to the asymptotically stable large-radius region. The necessary condition $r>3M$ alone is therefore not sufficient to determine the physical ISCO.

The motion of electrically charged particles was studied in an external test electromagnetic field. We obtained the corresponding radial equation, effective-potential branches, and circular-orbit angular momentum. Because algebraic manipulations involving squared circularity equations may introduce extraneous solutions, all candidate roots must be substituted back into the original unsquared equation and checked against the physical requirements of nonnegative angular-momentum squared and future-directed motion. These results describe charged-particle probes in a prescribed electromagnetic field and should not be interpreted as the dynamics of a self-consistently charged symmergent black hole solution.

Spinning particles were treated within the Mathisson--Papapetrou--Dixon formalism supplemented by the Tulczyjew spin supplementary condition. We obtained the radial momentum equation and the corresponding effective-potential branches. Since the four-momentum and tangent four-velocity are not generally parallel under the Tulczyjew condition, the physically relevant radial quantity is $(p^r/m)^2$ rather than $(u^r)^2$. Admissible spinning-particle trajectories must satisfy the discriminant and denominator constraints, possess a future-directed timelike tangent vector, and remain within the pole--dipole regime, $|s|/M\ll1$.

For photons emitted by circular geodesic sources, we derived the two signed observational frequency-shift branches and the identity
\begin{align}
(1+z_+)(1+z_-)=\frac{1}{A(r_e)}.
\end{align}
This identity reconstructs the lapse function at the emitter radius under the assumptions of the model. The approaching-side branch is a true blueshift only where $z_-<0$; it should therefore be displayed with its sign rather than through $|z_-|$. Moreover, disconnected radial bands produced by the oscillatory branch are bands of real circular-orbit candidates. Their stability must be verified separately through the second derivative of the effective potential.

All quantitative results are restricted to $|\epsilon f(r)|\leq0.1$. This requirement concerns the actual conformal deformation and cannot be replaced by the statement $|\gamma|\ll1$. We have also shown that a nontrivial decaying $\gamma>0$ mode cannot be simultaneously regular at the Schwarzschild horizon under the standard boundary assumptions. The Yukawa solution used here should therefore be interpreted as a perturbative exterior configuration over a restricted radial domain unless a separate near-horizon completion is supplied.

Within these limitations, the two signs of $\gamma$ lead to qualitatively distinguishable strong-field phenomenology. The Yukawa branch produces smooth, localized deviations, whereas the oscillatory branch can generate alternating radial structure in particle motion and photon frequency shifts. These signatures can provide complementary tests of the variable-curvature sector, although extracting the  boson-fermion imbalance requires independent control of the deformation amplitude, oscillatory phase, source radius, and astrophysical propagation effects.

\acknowledgments
B. P. and A. {\"O}. would like to acknowledge networking support by the COST Actions CA21106 - COSMIC WISPers in the Dark Universe: Theory, astrophysics and experiments (CosmicWISPers), CA23115 - Relativistic Quantum Information (RQI) and CA23130 - Bridging high and low energies in search of quantum gravity (BridgeQG) funded by COST (European Cooperation in Science and Technology). A. \"O. would also like to acknowledge networking support of the COST Action CA22113 - Fundamental challenges in theoretical physics (THEORY-CHALLENGES) and the COST Action CA21136 - Addressing observational tensions in cosmology with systematics and fundamental physics (CosmoVerse) funded by COST (European Cooperation in Science and Technology). They also thank to T\"UBITAK, ULAKBIM (T\"urkiye), and SCOAP$^3$ (Switzerland) for their support.

\bibliography{references}
\end{document}